\documentclass[reprint, superscriptaddress, amsmath, amssymb, aps]{revtex4-2}

\usepackage{graphicx}
\usepackage{dcolumn}
\usepackage{bm}
\usepackage{xcolor}
\usepackage{upgreek}
\usepackage[normalem]{ulem}
\usepackage[T1]{fontenc} 
\usepackage{braket} 
\usepackage{ragged2e} 
\usepackage{enumitem}
\usepackage[normalem]{ulem}
\usepackage{hyperref}
\usepackage{bibunits}
  
\hypersetup{urlcolor=blue, colorlinks=true, citecolor=blue, linkcolor=blue}

\defaultbibliographystyle{apsrev4-2}
\defaultbibliography{bibliography}

\makeatletter
\renewcommand{\figurename}{Fig.} 
\def\fnum@figure{\textbf{\figurename~\thefigure}}
\makeatother

\newcommand\LCN{\,London Centre for Nanotechnology, University College London, London WC1H 0AH, United Kingdom}
\newcommand\QMT{\,Quantum Motion, 9 Sterling Way, London N7 9HJ, United Kingdom}
\newcommand\CAM{\,Department of Materials Sciences and Metallurgy, University of Cambridge, Cambridge CB3 0FS, United Kingdom}

\newcommand\IMEC{\,IMEC, Kapeldreef 75, 3001 Leuven, Belgium}

\begin{document}
\begin{bibunit}
\preprint{APS/123-QED}

\title{Radiofrequency cascade readout of coupled spin qubits}

\author{Jacob F. Chittock-Wood} \thanks{Current address: Center for Emergent Matter Science, RIKEN, Saitama, Japan} \email{jacob.chittock-wood@riken.jp} \affiliation{\QMT} \affiliation{\LCN}
\author{Ross C. C. Leon} \affiliation{\QMT}
\author{Michael A. Fogarty} \affiliation{\QMT}
\author{Tara Murphy} \affiliation{\QMT}\affiliation{\CAM}
\author{Felix-Ekkehard von Horstig} \affiliation{\QMT}\affiliation{\CAM}
\author{Sofia M. Patom{\"a}ki} \affiliation{\QMT}\affiliation{\LCN}
\author{Giovanni A. Oakes} \affiliation{\QMT}
\author{James M. Williams} \affiliation{\QMT}\affiliation{\LCN}
\author{Nathan Johnson} \affiliation{\LCN}
\author{Julien Jussot}\affiliation{\IMEC}
\author{Stefan Kubicek}\affiliation{\IMEC}
\author{Bogdan Govoreanu}\affiliation{\IMEC}
\author{David F. Wise} \affiliation{\QMT}
\author{John J. L. Morton}\email{john@quantummotion.tech} \affiliation{\QMT}\affiliation{\LCN}
\author{M. Fernando Gonzalez-Zalba}
\thanks{Current address: CIC nanoGune Consolider, Donostia-San Sebastian, Spain \& IKERBASQUE, Basque Foundation for Science, Bilbao, Spain}
\email{fernando@quantummotion.tech}\affiliation{\QMT}

\begin{abstract}
\noindent Silicon spin qubits based on metal-oxide-semiconductor (MOS) technology are compatible with semiconductor manufacturing and offer a route to scalable quantum processing. However, spin readout typically relies on proximal charge sensors, which add architectural complexity and limit qubit connectivity. In situ dispersive readout techniques are more compact, which can alleviate these constraints, but exhibit limited sensitivity. Here we report a radiofrequency electron-cascade readout method that enhances the dispersive signal through alternating-current electron co-tunnelling. With this approach, we achieve an enhancement in signal-to-noise ratio of more than $35~$dB, leading to a minimum integration time of $7.6 \pm 0.2~\upmu$s. We demonstrate singlet-triplet readout of two-electron spins in a natural silicon planar MOS quantum dot array, and coherent spin control using the exchange interaction, which forms the basis for entangling gates. We find dephasing times of up to $500~$ns and a gate quality factor that exceeds 10.
\end{abstract}

\maketitle

\noindent 

\noindent Silicon-based quantum processors have  been used to demonstrate high-fidelity qubit initialisation~\cite{Yoneda2020, Takeda:2022}, measurement~\cite{Connors20, Oakes:2023, Takeda2024}, and single-\cite{Yoneda18, wu2025simultaneoushighfidelitysinglequbitgates} and two-qubit control~\cite{Xue:2022, Madzik:2022, Noiri:2022, Tanttu23, Steinacker24} in small-scale devices of up to two qubits, with fidelities exceeding the $99\%$ threshold required to implement quantum error correction~\cite{Fowler12}. Electron spin qubits in quantum dots (QDs)  have also been used for simple instances of quantum error correction in a 3-qubit array~\cite{Takeda:2022}, and operation  of a 6-qubit processor~\cite{Philips2022}.  Notably, such silicon-based quantum processors can be fabricated using industrial manufacturing techniques and integrated with cryogenic electronics~\cite{Pauka2021, Ruffino2022},  offering a promising route to scaled quantum computing~\cite{Gonzalez-Zalba:2021}.

Spin qubits based on silicon metal-oxide-semiconductor (MOS) technology ~\cite{Veldhorst15, Yang2020} and Si/SiGe heterostructures~\cite{Maune2012, Philips2022, Noiri:2022} are of particular interest for industrial manufacturing. The MOS approach shares similarities with modern silicon field-effect transistor (FET) manufacturing, allowing the formation of MOS QDs in both planar devices~\cite{Tanttu23, Veldhorst15, Yang2020, Fogarty2018, Jock:2018, Jock:2022}, and etched silicon structures such as nanowires~\cite{Maurand:2016, Urdampilleta2019, Oakes:2023} or finFETs~\cite{Camenzind22, Zwerver:2022}. While the latter constrains the qubit topology to 2xN bilinear QD arrays, the former offers easier scalability towards two-dimensional QD arrays~\cite{Sieu25}, which are essential for implementing quantum error correcting codes such as the surface code~\cite{Fowler12}. Single-qubit performance in MOS devices fabricated using semiconductor manufacturing lines has been demonstrated~\cite{Maurand:2016, Zwerver:2022}, but quantum processors require two-qubit interactions to operate.

To enable further scaling, methods that simplify the readout infrastructure  are required. The current standard for sensing --- the radiofrequency (rf) single-electron transistor --- provides high-fidelity readout~\cite{Connors20, Takeda2024} at the cost of occupying substantial space on the qubit chip, which limits qubit connectivity. Fast and compact dispersive rf measurement techniques~\cite{Pakkiam18, West:2019, Zheng:2019} reduce the readout footprint. However, \textit{in-situ} dispersive readout metrics have seen limited progress in recent years~\cite{West:2019}.
\begin{figure*}[t]
    \includegraphics{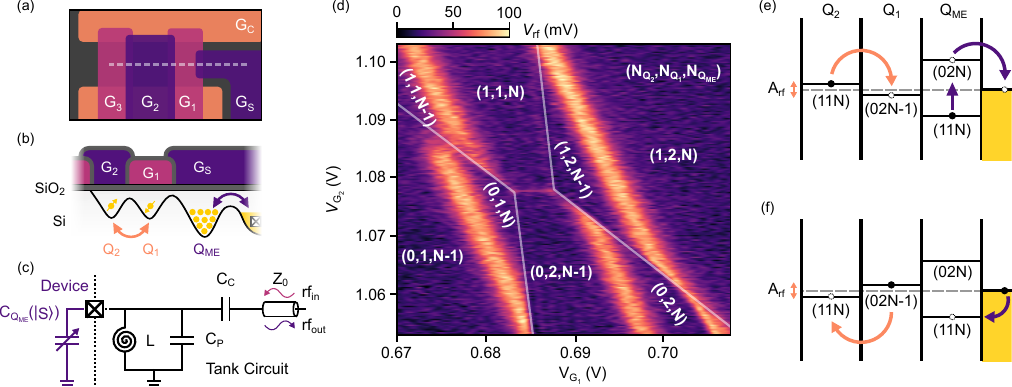}
    \def\figurename{Fig.}
    \caption{\textbf{$|$ Radio-frequency driven electron cascade}. 
    \textbf{a-b} Top view and cross-section (white-dashed line) schematic of the quantum dot (QD) array. Gates G$_1$ and G$_2$ define QDs, Q$_1$ and Q$_2$, tuned to the two-electron occupancy. The double quantum dot (DQD) is capacitively coupled to dot Q$_{\rm ME}$ occupied with many electrons and is controlled by gate G$_{\rm S}$. Arrows indicate single electron tunnelling events. \textbf{c} Schematic of the radiofrequency (rf) resonator bonded to the ohmic contact of the device, including a circuit equivalent representation of the QD array as a spin dependent variable capacitor $\rm C_{\text{Q}_{\text{ME}}}(\ket{S})$. The resonator is formed by an off-chip superconducting spiral inductor, $\rm L=136~$nH, arranged in parallel with the parasitic capacitance, $\text{C}_\text{P}=0.4~$pF, of the assembly. Connected to the transmission line $Z_0$, via coupling capacitor $\text{C}_{\rm C} = 0.1~$pF, rf$_{\rm in(out)}$ represents the incident (reflected) rf signal on the resonator. \textbf{d}, Charge stability diagram of the DQD as a function of gate voltages $V_{\rm G_1}$ and $V_{\rm G_2}$; $V_{\rm rf}$ denotes the demodulated rf voltage. \textbf{e-f}, Schematic representations of the cascade process in which an rf excitation with amplitude $\rm A_{rf}$ synchronously drives charge transitions within the QD array. The reservoir is shown as the shaded region; the dashed lines mark its Fermi level. \textbf{e}, A change in the charge occupation from ($\rm N_{{\rm Q}_2}$,$\rm N_{{\rm Q}_1}$,$\rm N_{{\rm Q}_{\rm ME}}$)=(1,1,N) to (0,2,N-1) raises the electrochemical potential of the Q$_\text{ME}$ above the Fermi level, causing one electron to synchronously escape to the reservoir. \textbf{f}, When the DQD is driven back to (1,1,N), an electron tunnels back from the reservoir to Q$_\text{ME}$.}
    \label{Fig:1}
\end{figure*}

In this Article, we report an \textit{in-situ} dispersive sensing mechanism that is termed \emph{radiofrequency electron cascade}  and can offer high spin qubit readout sensitivity.  With the readout method, we achieve a minimum integration time of $7.6\pm0.2~\upmu$s and, using a physical model~\cite{Gambetta07}, we calculate a 67\% singlet-triplet readout fidelity limited by spin relaxation. Second, we  also show control of an exchange-mediated coherent interaction, which forms the basis for a $\sqrt{\rm SWAP}$ gate between two spin qubits. 

\section*{Radiofrequency electron cascade readout} \label{sec:casc}
Reading out spin qubits within semiconductor QDs typically involves mapping a spin state of interest onto a charge state of one or more QDs~\cite{Ono2002, Elzerman2004}, which can then be detected using a variety of charge sensing methods~\cite{Vigneau:2023}. For example, Pauli spin blockade (PSB) can be used to map the singlet and triplet states of a pair of spin qubits onto two different charge configurations of a DQD (e.g.\ (1,1) or (0,2)), which are then detected by a single-electron transistor~\cite{Schoelkopf1998} or single-electron box~\cite{Oakes:2023}. \emph{In-situ} dispersive readout of a DQD combines these into a single step, using PSB to directly distinguish between singlet and triplet states through their difference in AC polarisability~\cite{Betz:2015}. This difference in polarisability is detected by incorporating the DQD into a radiofrequency tank circuit and measuring changes in the reflected rf signal. 
However, \textit{in-situ} dispersive readout has suffered from low sensitivity in planar MOS silicon quantum devices due to the relatively low gate lever arms~\cite{West:2019}. To improve the sensitivity of this technique, we introduce a third quantum dot (QD) coupled to a charge reservoir, which acts as an amplifier in measuring the AC-polarisability of the DQD. 
Instead of measuring the usual single-electron alternating current generated by cyclic tunneling between the two-spin singlet states of the DQD~\cite{Vigneau:2023}, we leverage the \emph{synchronised} single-electron AC current at the third dot-reservoir system generated as a consequence of the strong capacitive coupling to the DQD. Our approach offers the benefit of charge enhancement techniques such as latching~\cite{Harvey-Collard:2018}, DC cascading~\cite{vanDiepen:2021} and spin-polarized single-electron boxes~\cite{Urdampilleta2019} while retaining the non-demolition nature of \textit{in-situ} dispersive readout methods~\cite{Gusenkova2021}.

\begin{figure}[t] 
    \includegraphics[width = \linewidth]{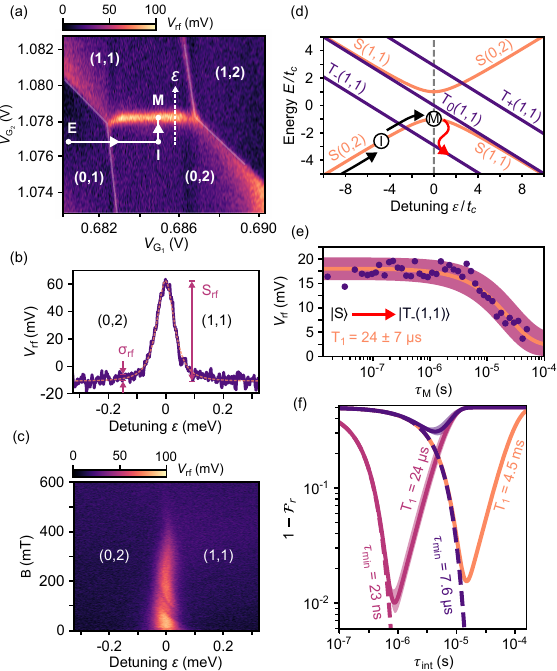}
    \def\figurename{Fig.}
    \caption{\textbf{$|$ Radiofrequency cascaded singlet-triplet readout.} \textbf{a} Charge stability diagram of the double quantum dot (DQD) around the (1,1)-(0,2) interdot charge transition (ICT), with a voltage pulse sequence (white arrows) and detuning axis $\varepsilon$ overlaid; $V_{\rm rf}$ denotes the demodulated rf voltage. \textbf{c} $V_{\rm rf}$ as a function of detuning across the ICT around $\varepsilon = 0$, fit to Eq.~(\ref{eqn:ICT-fit}) (dashed orange line). $S_{\rm rf}$ and $\sigma_{\rm rf}$ represent the signal amplitude and standard deviation respectively. \textbf{c}, Magneto-spectroscopy of the ICT as a function of applied magnetic $B$ and $\varepsilon$. \textbf{d} Energy diagram showing the dependence of the two-electron spin states, singlet $\ket{\text{S}}$ and triplet $\ket{\text{T}}$ as a function of $\varepsilon$ with respect to the interdot tunnel coupling $t_c$. The pulse sequence depicts the initialisation (I) of the $\ket{{\rm S(0,2)}}$ via an adiabatic ramp from the $(0,1)$ empty (E) state, followed by a non-adiabatic pulse to $\varepsilon = 0$ for measurement (M). \textbf{e}, $V_{\rm rf}$ as a function of wait time $\tau_{\rm M}$ at $\varepsilon = 0$ before measurement, following the pulse sequence depicted in panels \textbf{a} and \textbf{d}. The fitted line indicates a $\ket{\text{S}}$ to $\ket{{\rm T_-(1,1)}}$ relaxation time $T_1 = 24~\upmu$s. \textbf{f}, Calculated readout infidelity $1-\mathcal{F}_r$ as a function of integration time $\tau_{\text{int}}$ as described by Eq.~(\ref{eq:fidelity}). Solid lines depict relaxation time $T_1$ limited fidelity, whereas dashed lines depict $T_1 = \infty$. Parameters used from left to right include: $\tau_{\text{min}} = \left\{23~\text{ns}, 7.6~\upmu\text{s}, "\right\}$, $T_1 = \left\{24~\upmu\text{s}, ", 4.5~\text{ms}\right\}$. In both \textbf{e} and \textbf{f} shaded areas indicate the propagated error from the $\pm 1$ standard deviation of fit parameters. 
    } \label{Fig:2}
\end{figure}

\begin{figure*}[t]
    \includegraphics{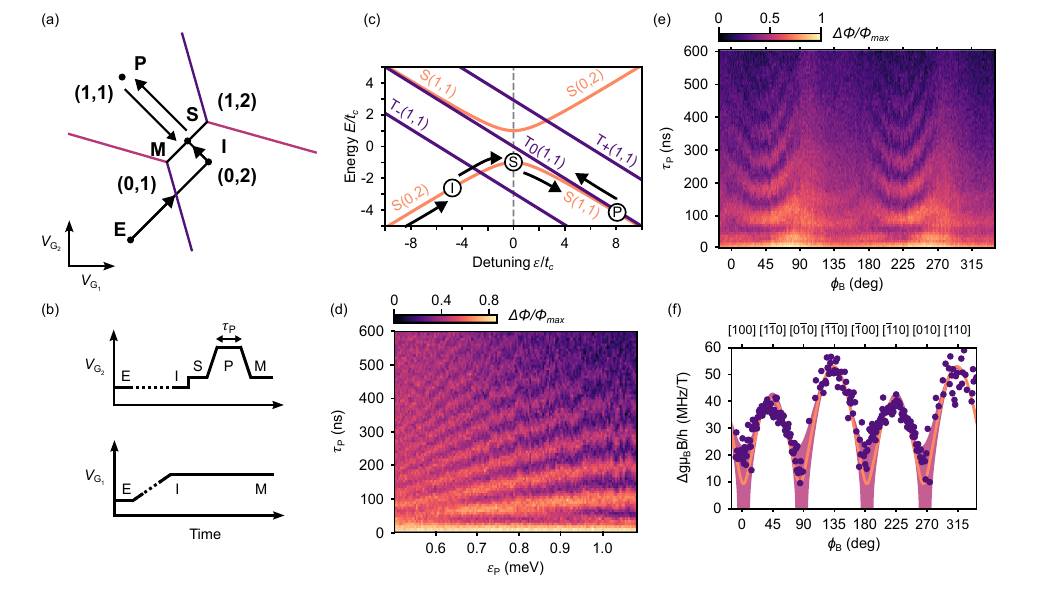}
    \def\figurename{Fig.}
    \caption{\textbf{$|$ Measurement of the spin-orbit interaction}. \textbf{a}, Schematic of the double quantum dot charge stability diagram around the (1,1)-(0,2) interdot charge transition with voltage pulse sequence EISPM overlaid (see Methods) \textbf{b}, The pulse sequence shown as a function of time where dashed lines indicate longer durations. \textbf{c}, Energy diagram showing dependence of the two-electron spin states, singlet $\ket{\text{S}}$ and triplet $\ket{\text{T}}$ as a function of detuning $\varepsilon$ with respect to the interdot tunnel coupling $t_c$. \textbf{d}, $\ket{\text{S}}$-$\ket{\text{T}_0}$ oscillations as a function of duration $\tau_P$ and detuning at point P, $\epsilon_P$, in the pulse sequence (applied magnetic field $B = 250$~mT, in-plane magnetic field orientation $\phi_B=235^{\circ}$). \textbf{e-f}, $\ket{\text{S}}$-$\ket{\text{T}_0}$ oscillation frequency dependence as a function of $\phi_B$, measured at fixed detuning $\varepsilon_P = 0.926\ $meV. The oscillation frequencies are obtained by Fourier analysis. \textbf{f}, The spin-orbit interaction component of the extracted frequencies, $\Delta g \mu_{\rm B} B/h$, are fit (line) using the model described in Eq.~(\ref{eqn:SOI}); where $\Delta g$ is the difference in g-factor between quantum dots, $\mu_B$ is the Bohr magneton, $B$ is the applied magnetic field, and $h$ is Planck's constant. The shaded area shows the propagated error from the $\pm1$ standard deviation of fit parameters.
    }
    \label{Fig:3}
\end{figure*}

\begin{figure*}[t]
    \includegraphics{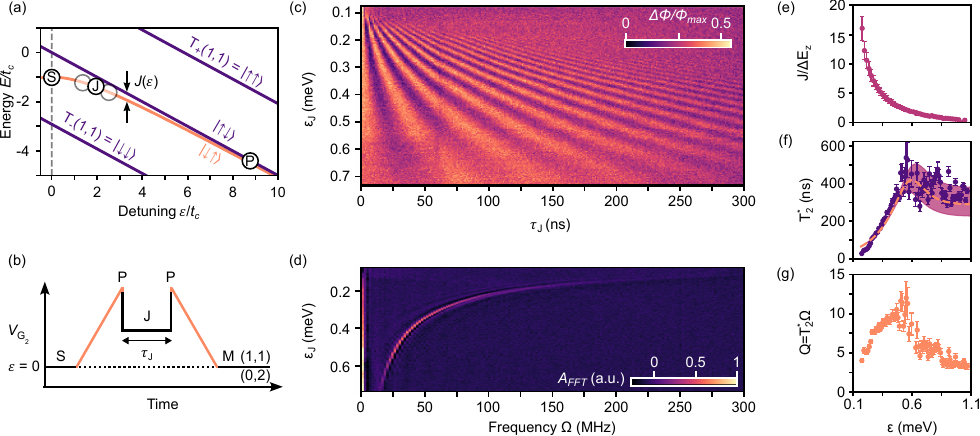}
    \def\figurename{Fig.}
    \caption{\textbf{$|$ Exchange control in a $^{\rm nat}$Si double quantum dot device}. \textbf{a}, Energy diagram depicting two-electron spin states in the (1,1) detuning $\varepsilon$ regime, with pulse sequence steps overlaid. Each axis is plotted with respect to the interdot tunnel coupling $t_c$. \textbf{b}, Detuning pulse sequence including initialisation (P) to the $\ket{\downarrow \uparrow}$ state via a semi-adiabatic ramp (orange), followed by a non-adiabatic pulse \text{(J)} to near zero-detuning to increase the exchange coupling $J(\varepsilon)$ for duration $\tau_J$. \textbf{c}, Exchange driven oscillations between $\ket{\downarrow \uparrow}$ and $\ket{\uparrow \downarrow}$ states, measured rf-phase response proportional to singlet probability, with \textbf{d}, the corresponding fast Fourier transform amplitude $A_{\rm FFT}$. \textbf{e} Ratio between exchange coupling strength $J(\varepsilon)$ and dot-to-dot Zeeman energy difference $\Delta E_z$ \textbf{f} Dephasing time $T_2^*$ (purple dots) extracted from the decay of the exchange oscillations and fit with $T_2^* = \sqrt{2}\hbar/\delta{\left(h \Omega\right)}$ (orange dashed line)~\cite{Dial:2013, Wu:2014}. The shaded area shows the propagated error from the $\pm1$ standard deviation of fit parameters. \textbf{g} Qubit quality factor $Q=T_2^* \Omega$. In \textbf{e-g} each data point is extracted as a fit parameter at a fixed detuning $\varepsilon_J$ point of panel \textbf{c}, with each error bar representing the standard deviation of a fitted parameter.
    }
    \label{Fig:4}
\end{figure*}

We use a device based on planar silicon MOS technology with an overlapping gate design~\cite{stuyck2021uniform} (see Fig~\ref{Fig:1}(a-b), Methods and Supplementary Section 1). Quantum dots Q$_1$ and Q$_2$ form a DQD which we tune to hold two electrons, while Q$_2$ is capacitively coupled to a multi-electron quantum dot (Q$_{\rm ME}$) that can exchange electrons with a charge reservoir. To measure the charge state of the system, we connect a superconducting spiral inductor to the reservoir, forming a $LC$ resonator (see Fig.~\ref{Fig:1}(c)). At voltages where the charge in Q$_{\rm ME}$ is bistable, cyclic tunneling generated by the small rf signal supplied to the resonator produces a change in capacitance that can be detected as a change in the phase response, $\Delta \Phi$, or demodulated voltage, $V_{\rm rf}$, using homodyne techniques~\cite{Vigneau:2023}. Lines in gate voltage space showing Q$_{\rm ME}$ charge bistability are shifted when intersecting charge transitions of the DQD, as shown in Fig.~\ref{Fig:1}(d). Such shifts form the basis of dispersive charge sensing measurements~\cite{Oakes:2023,Hogg:2023} which we do not exploit here. Instead, we focus on the directly observable AC signal in the region of gate voltage space where charge transitions may occur between Q$_1$ and Q$_2$. We ascribe this signal to a two-electron charge cascade effect driven by the rf excitation, which we explain using the diagrams in Fig.~\ref{Fig:1}(e-f). Consider an rf cycle in which the system starts in the occupation configuration ($\text{N}_{\text{Q}_1}$,$\text{N}_{\text{Q}_2}$,$\text{N}_{\text{Q}_{\text{ME}}})=(1,1,\text{N})$. Due to the strong capacitive coupling between Q$_2$ and Q$_{\rm ME}$, the rf excitation that drives the DQD from the $(1,1)$ to $(0,2)$ state synchronously forces an electron out of Q$_{\rm ME}$ in a cascaded manner, leading to $(0,2,\text{N}-1)$. The second half of the rf cycle then reverses the process. Overall, the rf cascade measures the polarizability of the DQD system, as for \textit{in-situ} dispersive readout measurements~\cite{West:2019, Zheng:2019}, but with the substantial advantage that the induced charge can be much greater, resulting in a dispersive measurement with greater sensitivity (see Supplementary Section 2 for more discussion on the requirements for rf electron cascade readout). Specifically, we find the power signal is amplified in the cascade approach compared to direct dispersive readout, by a factor
\begin{equation} \label{eqn:amp_factor}
	A= \left(1+\frac{1-\alpha_\text{R,ME}}{\alpha_\text{R,2}-\alpha_\text{R,1}}\right)^{2}>1,
\end{equation}

\noindent where $\alpha_{{\rm R},j}$ represents the gate lever arm between the reservoir and QD $j$. In particular, the sensor detects not only the interdot gate polarization charge $(\alpha_\text{R,1}-\alpha_\text{R,2})e$ but also the cascaded charge collected at the reservoir, $(1-\alpha_\text{R,ME})e$, (see Supplementary Section 3 for derivation). By comparing the measured signal-to-noise ratio (SNR) with and without the cascade effect, we find a lower bound for the power amplification factor, $A\geq(3.4\pm0.1)\times10^3$ $(+35.4~\text{\rm dB})$. The cascade SNR is extracted from the interdot charge transition fit shown in Fig.~\ref{Fig:2}(a-b) (see Eq.~(\ref{eqn:ICT-fit}-\ref{eqn:SNR}) in Methods). For the case where the cascade is absent, the SNR is assumed to be upper bounded by $<0.5$, as there is no observable signal, as shown in Extended Data Fig.~\ref{FigSI:CSD}(a) and (e). From the cascade SNR we extract a minimum integration time of $\tau_{\rm min} = 7.6\pm0.2~\upmu{\rm s}$ (see Eq.~(\ref{eqn:tmin}) in Methods), representing an improvement of over two orders-of-magnitude on prior planar MOS demonstrations of \emph{in-situ} dispersive readout~\cite{West:2019}.

We use the rf-cascade to distinguish between the singlet and triplet states of the DQD via PSB. The signature of PSB can be observed by measuring the asymmetric disappearance of the interdot charge transitions as a function of increasing the applied magnetic field~\cite{Betz:2015}, as shown Fig.~\ref{Fig:2}(c). At low magnetic field $(B\leq 200~\text{\rm mT})$, the system is free to oscillate between singlet states ($\ket{\text{S}(1,1)}\leftrightarrow\ket{\text{S}(0,2)}$) due to the action of the rf drive, yielding a signal in the rf response. However, at higher fields $(B\geq 200~\text{\rm mT})$, the polarised triplet $\ket{\text{T}_-(1,1)}$ state becomes the ground state for $\varepsilon \geq 0$ (as shown in Fig.~\ref{Fig:2}(d)), preventing a charge transition and resulting in the disappearance of the signal, initially for the region of the transition closer to the $(1,1)$ charge configuration. A quantum capacitance-based simulation of the data in Fig. \ref{Fig:2}(c) indicates an inter-dot tunnel coupling of $t_c=2.4$~GHz and electron temperature $T_e=50$~mK (see Supplementary Section 5). Figure~\ref{Fig:2}(e) shows the decay from $\ket{\text{S}}$ to $\ket{\text{T}_-(1,1)}$ at $\varepsilon = 0$ for applied magnetic field $B=250~$mT. The corresponding relaxation time is $T_1 = 24\pm7~\upmu$s, which is consistent with a study of a similar device, where a relaxation time of $T_1 = 10~\upmu{\rm s}$ was observed near zero detuning for a comparable tunnel coupling ($t_c\geq 1.9~$GHz)~\cite{Laine25}. In that study, $T_1$ was found to vary exponentially with $t_c$, with relaxation times exceeding $100~$ms for $t_c < 1~$GHz. Lowering $t_c$ would also enhance the readout sensitivity, as the dispersive signal is maximised at $2t_c =f_\text{rf}/2$~\cite{Peri25}, where $f_\text{rf}$ is the resonator drive frequency, which we set to 512 MHz.

We assess the performance of the radio-frequency driven electron cascade by calculating the readout fidelity, $\mathcal{F}_r$ as a function of integration time, using Eq.~(\ref{eq:fidelity}) in Methods~\cite{Gambetta07}. We calculate a relaxation-limited readout fidelity of $\mathcal{F}_r= 67\pm1\%$ based on the experimentally obtained $\tau_{\rm min} = 7.6~\upmu$s and integration time $\tau_{\text{int}} = 8~\upmu$s. The corresponding readout infidelity $1 - \mathcal{F}_r$ is depicted in Fig.~\ref{Fig:2}(f) (dark purple line). This infidelity is comparable to prior \emph{in-situ} readout demonstrations, but is achieved with integration times that are 1-2 orders of magnitude faster than those reported in previous silicon planar MOS and implanted donor systems~\cite{West:2019, Pakkiam18} (see Supplementary Section 7). To achieve a $99\%$ fidelity, two strategies are indicated in Fig.~\ref{Fig:2}(f): ($i$) reducing the minimum integration time to $24~$ns (light purple line), by enhancing the SNR through resonator optimisation, quantum-limited amplification~\cite{Vigneau:2023, Apostolidis24}, or both; and ($ii$) by extending the relaxation time of the system, which for the $T_1 = 4.5~$ms reported in Ref.~\cite{West:2019} yields the orange line.

We highlight that, unlike other charge enhancement techniques \cite{Harvey-Collard:2018, Urdampilleta2019, vanDiepen:2021}, the rf electron cascade retains the non-demolition nature of \textit{in-situ} dispersive readout measurements since the DQD system remains in an eigenstate after a measurement is performed~\cite{Nakajima2019, Yoneda2020}. We note that the rf excitation is continuously applied for all measurements in this Article, we discuss the potential impact of this in the echo sequence section.
\section*{Characterising spin-orbit coupling} \label{sec:spin-orbit}
Having established a method to distinguish between singlet and triplet spin states, our goal is to prepare and coherently control spin states of the DQD through voltage pulses along the detuning axis, $\epsilon$. Such pulses bring the DQD: i) from the (0,2) charge configuration in which a singlet is prepared; ii) into the (1,1) region where the electron spins are spatially separated between QDs and may evolve; and iii) back to an intermediate point where they can be measured (see Fig.~\ref{Fig:3}(a-c)). Deep in the (1,1) region, the spin basis states are predominantly $\ket{\uparrow \downarrow}$ and $\ket{\downarrow \uparrow}$. Under an adiabatic ramp to $\epsilon=0$ for readout, these two states map onto the $\ket{\text{T}_0(1,1)}$ and $\ket{\text{S}(0,2)}$, respectively. The basis states are separated in energy by $h \Omega = \sqrt{J(\varepsilon)^2 + \Delta E_z^2}$, where we include the kinetic exchange interaction $J(\varepsilon)$ and the Zeeman energy difference between electrons in each dot $\Delta E_z$. The spin detuning $\Delta E_z= \Delta g \mu_B B + g \mu_B\Delta B_\text{HF}$ (where $\mu_B$ is the Bohr magneton, $h$ is Planck's constant) contains two main contributions: (i) the difference in g-factor between QDs $\Delta g = \left|g_2 - g_1\right|$ arising from variations in the spin-orbit interaction (SOI) present near the Si/SiO$_2$ interface
~\cite{VeldhorstSOI2015, Tanttu:2019, Jock:2018}; and (ii) the difference in the effective $^{29}$Si nuclear magnetic field experienced by each QD, $\Delta B_\text{HF}$. The random fluctuations in the effective magnetic field experienced by each electron in the DQD can be described by a normal distribution with mean of 0 (given the negligible spin polarisation) and standard deviation $\sigma_\text{HF}=30\pm4~\upmu$T, as we shall see later. This value corresponds to a hyperfine energy strength of $3.4\pm0.4$~neV, which aligns well with other reports in natural silicon~\cite{Maune2012, Wu:2014, Connors22, Liles:2023}.

In prior work, the spin detuning $\Delta E_z$ has been leveraged to drive oscillations between $\ket{\text{S}}$ and $\ket{\text{T}_0}$ states \cite{Jock:2018, Jirovec:2021, Jock:2022, Liles:2023}. At an applied magnetic field $B = 250\ $mT, we observe similar oscillations using the pulse sequence presented in Fig.~\ref{Fig:3}(a-c): 
We start in the (0,1) configuration by emptying dot Q$_1$, then initialise the $\ket{\text{S}(0,2)}$ state via an adiabatic ramp across the (0,1)-(0,2) charge transition. A fast non-adiabatic pulse to $\varepsilon_P$ in the (1,1) region leads to oscillations between $\ket{\text{S}}$ and $\ket{\text{T}_0}$ over the course of the dwell time $\tau_P$. The final state is then measured dispersively using a non-adiabatic pulse back to the (1,1)-(0,2) charge transition at $\varepsilon=0$ for readout. The $\ket{\text{S}}$-$\ket{\text{T}_0}$ oscillations shown in Fig.~\ref{Fig:3}(d) provide a direct measurement of $\Omega$. For $\varepsilon_P \gtrsim 0.9\ $meV the dependence of the oscillation frequency on detuning is significantly reduced, suggesting that in this region the $\Delta E_z$ term dominates ($J(\epsilon) \leq \Delta E_z$), since $\Delta g$ is only weakly dependent on detuning $\left(\partial\Delta g/\partial\varepsilon \approx 0\right)$~\cite{Jirovec:2021, Jock:2022}. As we shall see later, at the deepest detuning ($\varepsilon_P= 1.054~$meV), we find $J/\Delta E_z=0.5\pm0.3$. 

The strength of the SOI which leads to the $\Delta g$ term depends on Rashba and Dresselhaus spin-orbit couplings. The SOI (and hence $\Delta g$) can be tuned by varying the electrostatic confinement perpendicular to the interface and the transverse magnetic field \cite{VeldhorstSOI2015, Jock:2018}. We vary the orientation of the in-plane magnetic field and observe changes in the $\ket{\text{S}}$-$\ket{\text{T}_0}$ oscillation frequency, as shown in Fig.~\ref{Fig:3}(e-f). We fit the variation in $\Omega$ as a function of the angle, $\phi_B$, between the [100] crystal axis and the applied (in-plane) magnetic field \cite{Jock:2018},

\begin{equation} \label{eqn:SOI}
    \Delta E_z/h = \sqrt{{\left(\left|B\right|\left|\Delta \alpha - \Delta \beta \sin(2 \phi_B)\right|\right)}^2 + {\left(g\mu_B\sigma_\text{HF}/h\right)}^2}.
\end{equation}
The Rashba and Dresselhaus SOI terms are respectively captured by $\Delta \alpha$ and $\Delta \beta$. We find $\Delta \alpha = 6.2\substack{+1.6 \\ -1.5}\ $MHz/T and $\Delta \beta = 45\substack{+5 \\ -7}~$MHz/T, which are larger than other reported values~\cite{Jock:2018, Tanttu:2019, Cifuentes:2023} and could be partially influenced by the large asymmetry in the gate biasing conditions. 
The fit assumes that for the fixed detuning of $\varepsilon_P=0.926\ $meV used here, the residual exchange interaction $J(\varepsilon_P)/h = 6.3\pm1.9\ $MHz is independent of in-plane magnetic field orientation. We operate at an in-plane magnetic field direction near the $\left[1\bar{1}0\right]$ direction at $\phi_B = 55^{\circ}~(235^{\circ})$. Overall, this section expands the recent studies of the SOI in isotopically purified $^{28}$Si MOS nanostructures~\cite{Tanttu:2019, Jock:2018, Jock:2022, Cifuentes:2023} to natural silicon, where the non-negligible effect of the Overhauser field needs to be taken into account. 

\section*{Exchange control} \label{sec:exchange}
We implement exchange control using the sequence depicted in Fig.~\ref{Fig:4}(a,b), where the $\ket{\downarrow \uparrow}$ state is initialised via a ramp from $\varepsilon = 0$ into the (1,1) configuration that is adiabatic with respect to $E_z$~\cite{Maune2012}. A fast non-adiabatic pulse towards zero detuning increases the exchange coupling, driving oscillations between the $\ket{\downarrow \uparrow}$ and $\ket{\uparrow \downarrow}$ states at frequency $\Omega(\varepsilon)$, as observed in Fig.~\ref{Fig:4}(c). The final state after some evolution time $\tau_J$ is projected to $\ket{\text{S}}$ or $\ket{\text{T}_0}$ for readout. The Fourier transform of the exchange oscillations (see Fig.~\ref{Fig:4}(d)) reveals a single peak of increasing frequency as the detuning is reduced, indicating the purity of the oscillations and the enhanced exchange strength at lower detuning. From this, we observe that the exchange coupling is tunable over a range of $J=5 - 122~$MHz.

To quantify the properties of these rotations, we combine the results of the exchange oscillations in Fig.~\ref{Fig:4}(c) and the $\ket{\text{S}}$-$\ket{\text{T}_0}$ oscillations in Fig.~\ref{Fig:3}(d), to extract the ratio $J/\Delta E_z$ and the intrinsic coherence time $T_2^*$ over a wide range of detunings, see Fig.~\ref{Fig:4}(e,f). We extract $T_2^*$ by fitting the oscillations at each detuning point with a Gaussian decay envelope of the form exp$\left[-(\tau/T_2^*)^2\right]$, and then obtain $\Delta E_z = 9.6\pm1.2~$MHz from the fit to the expression 

\begin{equation}\label{eq:t2}
    \frac{1}{T_2^*}=\frac{1}{\sqrt{2}\hbar}\sqrt{\left(\frac{J}{h\Omega}\frac{dJ}{d\varepsilon}\delta\varepsilon_\text{rms}\right)^2+\left(\frac{\Delta E_z}{h\Omega}\delta\Delta E_{z,\text{rms}}\right)^2}, 
\end{equation}

\noindent where $\delta\varepsilon_\text{rms}$ and $\delta\Delta E_{z,\text{rms}}$ refer to the root mean square of the fluctuations in $\varepsilon$ and $\Delta E_{z}$~\cite{Dial:2013, Wu:2014}.

The extracted $J/\Delta E_z$ ratio is shown in Fig.~\ref{Fig:4}(e), reducing as a function of increasing $\varepsilon$ to a minimum value of $0.5\pm0.3$ at $\varepsilon = 1.054$ meV (beyond this point we cease to observe oscillations). This non-zero minimum shows there remains a residual exchange that cannot be fully turned off, which should be taken into account when designing two-qubit exchange gates.

From the $T_2^*$ data shown in Fig.~\ref{Fig:4}(f), we observe a rapid increase in coherence as the detuning increases from zero, indicative of a low $\delta\varepsilon_\text{rms}$. The extracted value of $\delta \varepsilon_{rms} = 5.4\pm0.1~\upmu$eV, obtained over a measurement time of 0.3~h per trace, is at the state-of-the-art for MOS devices~\cite{Jock:2018, Fogarty2018, Liles:2023}, and can be attributed to the low charge noise achieved for samples using this 300 mm process~\cite{Elsayed:2024}, comparable to reports on SiGe heterostructures~\cite{PaqueletWuetz23} (see Methods). As the detuning increases further, where $J < \Delta E_z$, we observe that noise in $\Delta E_z$ dominates (due to $^{29}$Si nuclear spins), leading to a relatively constant $T_2^*$. From this saturation value of $T_2^*=0.28\pm0.04~\upmu$s, we extract $\sigma_\text{HF}=\sqrt{2}\pi\delta\Delta E_\text{z,rms}/(g\mu_B)=30\pm4~\upmu{\rm T}$. Note that we assume the Zeeman energy fluctuations are dominated by the Overhauser field rather than noise in the g factor difference.

The entangling two-qubit gate achieved between the spin qubits under the exchange interaction depends on the ratio $J/\Delta E_z$, tending to a $\sqrt{\text{SWAP}}$ operation as $J\gg\Delta E_z$ or a C-PHASE when $J\ll\Delta E_z$, though any gate within this set parameterised by $J/\Delta E_z$ can be used as the building block for a quantum error correcting code such as the surface code~\cite{Patomaki2024}. Defining the qubit quality factor as $Q = T_2^* \Omega$ (i.e.\ the number of periods before the amplitude of oscillations decays by $1/e$), we find $Q\gtrapprox 10$ in the region $J/\Delta E_z = 2.1 - 3.2$, on par with prior reports across a range of semiconductor platforms~\cite{Jock:2018, Dial:2013, Wu:2014, Jirovec:2021, Connors22} (see Supplementary Section 8). This provides an upper bound estimate on the achievable two-qubit gate fidelity using the approximation $\mathcal{F} \approx 1 - 1/4Q \lessapprox 98\%$ \cite{Stano:2022}. To implement error-correctable two-qubit gates this fidelity would need to surpass $99\%$~\cite{Fowler12}, which could be achieved using isotopically enriched silicon. In the next section, we extend the coherence time using spin refocusing techniques.

\begin{figure}[t] 
    \includegraphics{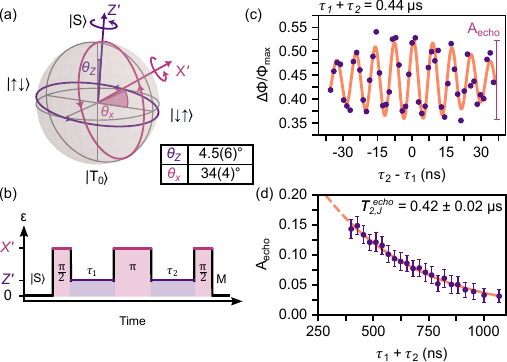}
    \def\figurename{Fig.}
    \caption{\textbf{$|$ Echo sequence}. \textbf{a}, Bloch sphere representation of the odd-parity two spin sub-space, indicating the rotation axes, $\hat{Z'}$ and $\hat{X'}$ and their angular deviation $(\theta_Z, \theta_X)$ from the nominal $\hat{Z}$ axis defined by $\ket{S}$ and $\ket{T_0}$. \textbf{b}, Schematic of the exchange echo sequence. \textbf{c}, Echo signal data (purple dots) as a function of free evolution time difference $\tau_2-\tau_1$, fit with a decay envelope (orange line) with amplitude $A_{\rm echo}$. \textbf{d}, Echo amplitude as a function of total free evolution time $\tau_1+\tau_2$ where $\tau_2=\tau_1$. The data (purple dots) were fitted with an exponential decay (dashed orange line). Error bars indicate the standard deviation of the fitted $A_{\rm echo}$ data.
    } \label{Fig:5}
\end{figure}

\section*{Echo Sequence} \label{sec:echo}

\begin{figure*}[t] 
    \includegraphics[width = \linewidth]{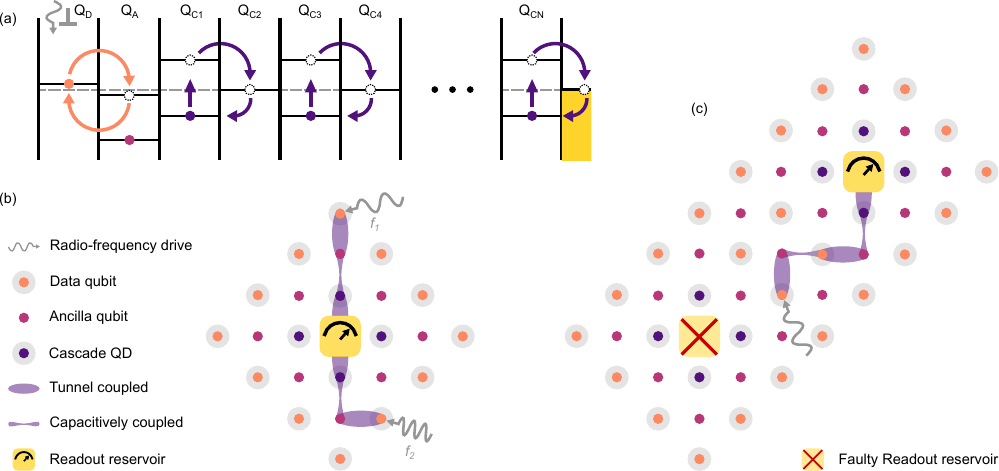}
    \def\figurename{Fig.}
    \caption{\textbf{$|$ Extending radio-frequency cascade readout to two-dimensional arrays}. \textbf{a}, Schematic representation of the cascade process extended to an arbitrarily long one-dimensional array. Pauli spin blockade readout is achieved by tunnel coupling the data Q$_{\rm D}$ (orange) and ancilla Q$_{\rm A}$ qubit (light purple), the resulting DQD is capacitively coupled to a chain of DQDs tuned into the cascade configuration (Q$_{\rm CN}$) (dark purple). Enabling in-situ dispersive readout at an arbitrary distance. \textbf{b}, Two-dimensional array schematic unit-cell centred around a readout reservoir (yellow), connected to a radio-frequency reflectometry readout apparatus. Simultaneous multiplexed readout is achieved by applying distinct rf frequencies $f_1$ and $f_2$ to separate electrostatic gates, each controlling data qubit QDs, driving distinct cascaded charge transitions. \textbf{c} Repetition of the unit cell highlighting the robustness of the scheme to points of failure, such as a faulty readout reservoir, by re-routing the readout cascade chain to the nearest reservoir. 
    } \label{Fig:6}
\end{figure*}

Dephasing of the two-electron spin state due to low-frequency electric or magnetic noise can be corrected using refocusing pulses. We implement an echo sequence by combining periods of evolution at different detuning points in order to achieve rotations around the two axes, $Z'$ and $X'$ shown in Fig.~\ref{Fig:5}(a). The specific sequence shown in Fig.~\ref{Fig:5}(b), termed the exchange echo, primarily reduces the impact of electric noise \cite{Dial:2013, Jock:2018}.

In the exchange echo sequence, after initialising a $\ket{\text{S}(1,1)}$ state and applying a $X'_{\pi/2}$ rotation, the two-electron system dephases under the effect of charge noise for a time $\tau_1$. The free evolution occurs at a detuning point where $J/\Delta E_z=12.8\pm1.6$ where we measured $T_{2}^* = 43\pm3~$ns (see Fig~\ref{Fig:4}(f)). We then refocus the spins by applying a $X'_\pi$ rotation and let the system evolve for $\tau_2$ until a second $X'_{\pi/2}$ rotation maps the resulting state to the $\ket{\text{S}}$-$\ket{\text{T}_0}$ axis. We extract the amplitude of the echo by fitting the signal in the $\tau_2-\tau_1$ domain (see Fig.~\ref{Fig:5}(c)), and plot its value as a function of total free evolution time $\tau_1+\tau_2$. Fig.~\ref{Fig:5}(d) shows the echo amplitude decays exponentially with total time ($\tau_2+\tau_1$), yielding a characteristic $T_{2,J}^\text{echo} = 0.42\pm0.02~\upmu$s which corresponds to an order of magnitude increase in the coherence time, similar to prior work with silicon devices~\cite{Jock:2018, Connors22}. From our fit to Eq.~\eqref{eq:t2}, we extract a magnetic-noise-limited $T_{2,\Delta E_z}^* = 3.3~\upmu$s at this set point, indicating that sources other than magnetic noise limit $T_{2,J}^\text{echo}$. This finding is in agreement with previous reports, which show that $T_{2,J}^{\rm echo}$ is an order of magnitude shorter than the magnetic noise limit~\cite{Jock:2018, Connors22} (see Supplementary Section 9). We speculate that residual high-frequency charge noise limits the echo coherence~\cite{Yoneda18}. This noise may include factors such as the effect of the rf readout excitation, which is active throughout the control sequence. We estimate the upper-bound on detuning fluctuations associated with this rf excitation to be $\delta\varepsilon_{\rm rms}^{\rm rf} \leq 2.7~\upmu{\rm eV}$ (see Methods Eq.~(\ref{eq:de_rf})). However, this limitation is not intrinsic to the the readout scheme itself, as the effect can be effectively mitigated by implementing a pulsed readout protocol. In this approach, the rf-drive is deactivated during the control sequence, and only activated during the readout phase.

\section*{Conclusions} \label{sec:disc}

We have reported the radiofrequency electron cascade as high gain in-situ dispersive readout technique, demonstrated here in planar MOS device. The radiofrequency driven electron cascade could be extended to larger arrays for distant readout, which would eliminate the need for swap-based schemes that rely on shuttling information to a sensor~\cite{Sigillito2019}. Such an extension builds upon existing schemes for two-dimensional grids using data and ancilla qubits (Fig.~\ref{Fig:6})~\cite{Fowler12, Oakes:2023}. We assume that the tunnel barriers between each QD can be precisely controlled, enabling optimal readout fidelity through tuning $t_c$~\cite{Takeda2024}. 

Figure~\ref{Fig:6}(a) illustrates the readout protocol in a simplified one-dimensional array. In order to readout the data qubit Q$_{\rm D}$ at the periphery of the unit cell, the data qubit is tunnel-coupled to a neighbouring ancilla qubit Q$_{\rm A}$. The ancilla qubit, in turn, is capacitively coupled to a chain of DQDs configured in the cascade configuration, eventually connected to a reservoir. An rf-drive applied to the ancilla-data DQD initiates the cascade, propagating along the chain to a readout reservoir. This scheme uniquely enables the simultaneous readout of distant data qubits through frequency multiplexing at the readout tank circuit. By driving multiple cascade chains at distinct frequencies, multiple distant qubits can be readout simultaneously (Fig.~\ref{Fig:6}b), in contrast to earlier schemes~\cite{vanDiepen:2021}. Furthermore, combining unit cells into a dense, scalable 2D grid allows readout resource sharing, resulting in the system being resilient to failure points such as faulty QDs or reservoirs (Fig.~\ref{Fig:6}c).

Furthermore, we have demonstrated exchange control, which forms the basis for two-qubit gates between spin qubits, in a natural planar silicon MOS DQD device. The measured detuning noise is state-of-the-art for planar MOS~\cite{Elsayed:2024}, and the $T_2^*$ is a relatively long for a natural silicon device~\cite{Stano:2022}. These results are consistent with reports on devices from the same 300~mm fabrication line~\cite{Elsayed:2024}, indicating that process quality is a contributing factor. Follow-up studies could include measurements in isotopically enriched Si samples (see Supplementary Section 10  and Ref.~\cite{Steinacker24}). Further work could also extend exchange control to larger spin qubit arrays, adapting dedicated gates~\cite{Yang2020, Tanttu23} to primarily control exchange strength over a wider range. This could enable symmetric exchange pulses and reduce $J/\Delta E_z$ well below 1, which should lead to an overall reduction in sensitivity to charge noise.

\section*{Methods} \label{Methods:1}
\textbf{Fabrication details.} The device measured in this study is fabricated on a natural silicon 300~mm wafer, with three 30~nm-thick in-situ $n^{+}$~phosphorus-doped polycrystalline silicon gate layers formed with a wafer-level electron-beam patterning process~\cite{stuyck2021uniform, Elsayed:2024}. We use a high-resistivity ($>$3 k$\Omega/$cm) $p$-type Si wafer. First, a $8$~nm-thick, high quality SiO$_{2}$ layer is grown thermally to minimize the density of defects in the oxide and at the interface. Then, we subsequently pattern the gate layers using litho-etch processes and electrically isolate them from one another with a 5~nm-thick blocking high-temperature deposited SiO$_{2}$~\cite{stuyck2021uniform}. We employ the first layer of gates (closest to the silicon substrate) to provide in-plane lateral confinement in the direction perpendicular to the double quantum dot axis. We use the second layer of gates (G$_2$ in this case) to form and control primarily quantum dot, Q$_2$. Finally, we use the third gate layer to form and control quantum dot, Q$_1$, via G$_1$ and both the multi-electron quantum dot, Q$_\text{ME}$, and reservoir via G$_\text{S}$, as shown in Fig.~\ref{Fig:1}(a-b). Supplementary Section 1 includes scanning and transmission electron microscopy images of structures similar to the quantum dot array.

\textbf{Measurement set-up.} We perform the measurements at the base temperature of a dilution refrigerator ($T\sim 10$\,mK). We send low-frequency signals through cryogenic low-pass filters with a cut-off frequency of 65~kHz, while we apply pulsed signals through attenuated coaxial lines. Both signals are combined through bias-Ts at the sample PCB (printed circuit-board) level. The PCB was made from RO4003C 0.8\,mm thick with an immersion silver finish. For readout, we use radio-frequency reflectometry applied on the ohmic contact of the device. We send radio-frequency signals through attenuated coaxial lines to an on-PCB $LC$ resonator, arranged in a parallel configuration, formed by a coupling capacitor ($C_\text{c}$), a 100~nm-thick NbTi superconducting spiral inductor ($L$) and the parasitic capacitance to ground ($C_\text{p}$), as shown in Fig.~\ref{Fig:1}(c). We drive the resonator at 512.25~MHz which is the frequency of the system when G$_\text{S}$ is well above threshold. The reflected rf signal is then amplified at 4\,K and room temperature, followed by quadrature demodulation, from which the amplitude and phase of the reflected signal were obtained (homodyne detection). 

\textbf{Radio-frequency cascade readout performance.} 
The SNR of the interdot charge transition shown in Fig.~\ref{Fig:2}(b) is determined by fitting the signal to an expression proportional to the quantum capacitance $\propto \partial^2 E/\partial \varepsilon^2$~\cite{Betz:2015, Vigneau:2023},
\begin{equation} \label{eqn:ICT-fit}
    V_{\text{I,Q}} \propto V_{\text{I,Q}}^{0} - S_\text{I,Q}\frac{(2t_c)^3}{\left((\varepsilon-\varepsilon_0)^2 + (2t_c)^2\right)^{\frac{3}{2}}},
\end{equation}
where $t_c$ is the tunnel coupling, $\varepsilon_0$ is the detuning value of the centre of the peak, and $S_\text{I,Q}$ and $V_{\text{I,Q}}^{0}$ are the voltage signal and voltage offset in phase and quadrature components of the measured rf-response, respectively. The power SNR is determined by combining the measured in-phase an quadrature voltage signal components of the reflected radio-frequency signal, 
\begin{equation} \label{eqn:SNR}
    \text{SNR} = \frac{S_I^2 + S_Q^2}{0.5(\sigma_I^2 + \sigma_Q^2)}
\end{equation}
where $S_{{\rm rf,I(Q)}}$ and $\sigma_{\rm rf,I(Q)}$ are indicated in Fig.~\ref{Fig:2}(b), representing the signal and the standard deviation in background signal (away from zero detuning), respectively. Here, we assume a white noise spectrum dominated by the first cryogenic amplification stage~\cite{vonHorstig23}. In the measurements acquired throughout the text, the applied rf-power is $P_{\text{rf}} = -88.5~$dBm, corresponding to a power $\text{SNR} = (1.7\pm0.1)\times10^3$. The minimum integration time is determined using~\cite{Vigneau:2023},
\begin{equation} \label{eqn:tmin}
    \tau_{\text{min}} = \frac{N_{\text{avg}}}{2\Delta f~\text{SNR}}
\end{equation}
where $N_{\text{avg}} = 4000$ is the number of averages used in the measurement, and $\Delta f = 1.53 f_{\text{LPF}}$ is the measurement bandwidth set by the low-pass filter cutoff frequency $f_{\text{LPF}} = 100~$kHz. Together, these two parameters give the noise equivalent integration time $\tau_{\rm NE} = 13~{\rm ms}$ for the measurement in Fig.~\ref{Fig:2}(b). Substituting the SNR into Eq.~(\ref{eqn:tmin}) we find a minimum integration time $\tau_{\text{min}} = 7.6\pm0.2~\upmu$s. In combination with the relaxation time $T_1$ extracted in Fig.~\ref{Fig:2}(e), the readout fidelity can be calculated using the following expression~\cite{Gambetta07},
\begin{equation} \label{eq:fidelity}
    \mathcal{F}_r = \frac{1}{2}\left[1 + \text{erf}\left(\sqrt{\frac{\tau_{\text{int}}}{8 \tau_{\text{min}}}}\right)\exp\left(-\frac{\tau_{\text{int}}}{2T_1}\right)\right]
\end{equation}
where $\tau_{\rm int}$ is the integration time and $\text{erf}(x)$ is the Gauss error function. The corresponding readout infidelity $1 - \mathcal{F}_r$ is shown in Fig.~\ref{Fig:2}(f).

We define the detuning fluctuations associated with the rf excitation to be given by~\cite{Vigneau:2023},
\begin{equation} \label{eq:de_rf}
    \delta\varepsilon_{\rm rms}^{\rm rf} = \frac{\alpha_{21} V_{\rm dev}}{\sqrt{2}},
\end{equation}
where $\alpha_{21} = \alpha_{\rm R,2}-\alpha_{\rm R,1}$ and $V_{\rm dev}$ is the voltage excitation at the reservoir ohmic contact, from input rf signal $V_{\rm in}$~\cite{Ahmed18},
\begin{equation}
    V_{\rm dev} = \frac{2C_CQ_LV_{\rm in}}{C_C +C_P}
\end{equation}
where $Q_L = 100$ is the loaded quality factor of the resonator depicted in Fig.~\ref{Fig:1}(c), and rms $V_{\rm in} = 8.4~\upmu{\rm V}$ corresponding to input $P_{\rm rf} = -88.5~{\rm dBm}$. We then estimate an upper bound to $\alpha_{21}\geq0.01$ by re-arranging Eq.~(\ref{eqn:amp_factor}), assuming $\alpha_{\rm R,ME} = 0.35\pm0.15$. The resulting rf-induced fluctuation in detuning is then, $\delta\varepsilon_{\rm rms}^{\rm rf} \leq 2.7 \pm 0.6~\upmu{\rm eV}$.

\textbf{Control sequence and charge noise estimation.} The pulse sequence implemented to demonstrate exchange control is as follows:
\begin{enumerate}[leftmargin=+.2in]
    \item E: Empty quantum dot $\text{Q}_1$ by biasing the gate voltages such that the ground state is in the (0,1) charge configuration, over a duration of 100 ns.
    \item E-I: Initialise the ground state $\ket{\text{S}}$ via an adiabatic ramp from the (0,1) to (0,2) configuration, over a duration of $10~\upmu$s.
    \item I-S: Non-adiabatic pulse across the $\ket{\text{S}}$-$\ket{\text{T}_-}$ anti-crossing to $\varepsilon = 0$, over a duration of $100~$ns. Establishing a symmetric detuning point to apply subsequent ramps.
    \item S-P: Adiabatic ramp with respect to $\Delta E_z$ to $\varepsilon = 0.926~$meV over a duration of 250 ns. Initialising $\ket{\downarrow\uparrow}$ where $J<\Delta E_z$.
    \item P-J-P: Non-adiabatic pulse to near zero detuning $\varepsilon_J$ and back where $J\gg\Delta E_z$, for variable duration $\tau_J$.
    \item P-M: Reverse adiabatic ramp to $\varepsilon = 0$, projecting the final state onto $\ket{\text{S}}$ or $\ket{\text{T}_0}$ for readout.
    \item M: Pre-measure delay of $6~\upmu$s is waited over before integrating for a measurement time of $8~\upmu$s.
\end{enumerate}
The total duty cycle of the control sequence, $T_{\text{rep}} \approx 25~\upmu$s, provides the high frequency $f_{\text{high}}$ bound we integrate over to estimate $S_0$, the power spectral density noise at $1~$Hz. In combination with the total time to acquire a trace of exchange oscillations, $T_M = 1/f_{\text{low}}=0.3~$h, we can estimate $S_0$ from \cite{Kranz2020},
\begin{equation}
\delta\varepsilon_{\text{rms}} = \sqrt{2\int_{f_{\text{low}}}^{f_{\text{high}}}{S_\varepsilon(f) df}}
\end{equation}
where the power spectral density of charge noise has the functional form $S_{\varepsilon}(f) = S_0/f^{\alpha}$ with typical values of $\alpha$ ranging from $1$ to $2$. Considering the measured integrated charge noise ($\delta\varepsilon_{rms} = 5.4~\upmu$eV), we estimate an upper and lower bound of $\sqrt{S_0^{\alpha = 1}} = 0.91~\upmu$eV$/\sqrt{\text{Hz}}$ and $\sqrt{S_0^{\alpha = 2}} = 0.12~\upmu$eV$/\sqrt{\text{Hz}}$ respectively. This result is on par with the state-of-the-art reported in planar MOS devices~\cite{Elsayed:2024} and strained Ge wells (Ge/SiGe)~\cite{Lodari2021} $\sqrt{S_0} = 0.6 \pm 0.3~\upmu\text{eV}/\sqrt{\rm Hz}$, as well as Si wells in Si/SiGe heterostructures $\sqrt{S_0} = 0.3 - 0.8~\upmu{\rm eV}/\sqrt{\rm Hz}$~\cite{Connors22, PaqueletWuetz23}.

\section*{Data Availability}
\noindent The data that support the plots within this paper and other findings of this study are available from the corresponding authors upon reasonable request.

\section*{Acknowledgements}
\noindent We acknowledge helpful conversations with H.~Jnane, A.~Siegel, S.C.~Benjamin, A.J. Fisher and G. Burkard at Quantum Motion. We also acknowledge technical support from G.~Antilen Jacob at the London Centre for Nanotechnology. This work received support from the European Union's Horizon 2020 research and innovation programme under grant agreement No. 951852 (Quantum Large Scale Integration in Silicon); from the Engineering and Physical Sciences Research Council (EPSRC) under grant Nos.~(EP/S021582/1), (EP/L015978/1), (EP/T001062/1) and (EP/L015242/1); and from Innovate UK under grant Nos. (43942) and (10015036). T.M.~acknowledges support from the Winton Programme for the Physics of Sustainability. M.F.G.Z.~acknowledges support from the UKRI Future Leaders Fellowship (MR/V023284/1).

\section*{Author contributions}
\noindent J.F.C.W. conducted the experiments and analysed the results presented in this work with input from R.C.C.L., M.A.F., M.F.G.Z. and J.J.L.M.; J.F.C.W., M.A.F. and R.C.C.L. conducted preliminary experiments; T.M. and G.A.O developed the quantum dot rf simulator in Supplementary Section 4 informed by electrostatic modelling from J.W.; T. M. performed and wrote about the simulations under the supervision of D. F. W. and M. F. G. Z.; F.E.v.H. characterised the superconducting resonator, and carried out experiments on additional devices to validate the rf-cascade technique with support from J.F.C.W and M.F.G.Z.; S.M.P. designed the device under the supervision of M.A.F., M.F.G.Z. and J.J.L.M.; J.J. and S.K. fabricated the device under the supervision of B.G.; N.J. maintained the experimental setup; J.F.C.W., M.F.G.Z. and J.J.L.M. wrote the manuscript with input from R.C.C.L., M.A.F., N.J. and S.M.P.; M.F.G.Z. and J.J.L.M. conceived and oversaw the experiment.

\section*{Competing Interests}
\noindent Quantum Motion Technologies Limited has filed a patent application related to this work: EP23206749.6 (Europe) by J.F.C.W., R.C.C.L., M.A.F. and M.F.G.Z.

\putbib
\end{bibunit}


\newpage
\counterwithin{figure}{section}
\setcounter{figure}{0}
\renewcommand{\thefigure}{\arabic{figure}}
\begin{figure*}[t]
    \includegraphics{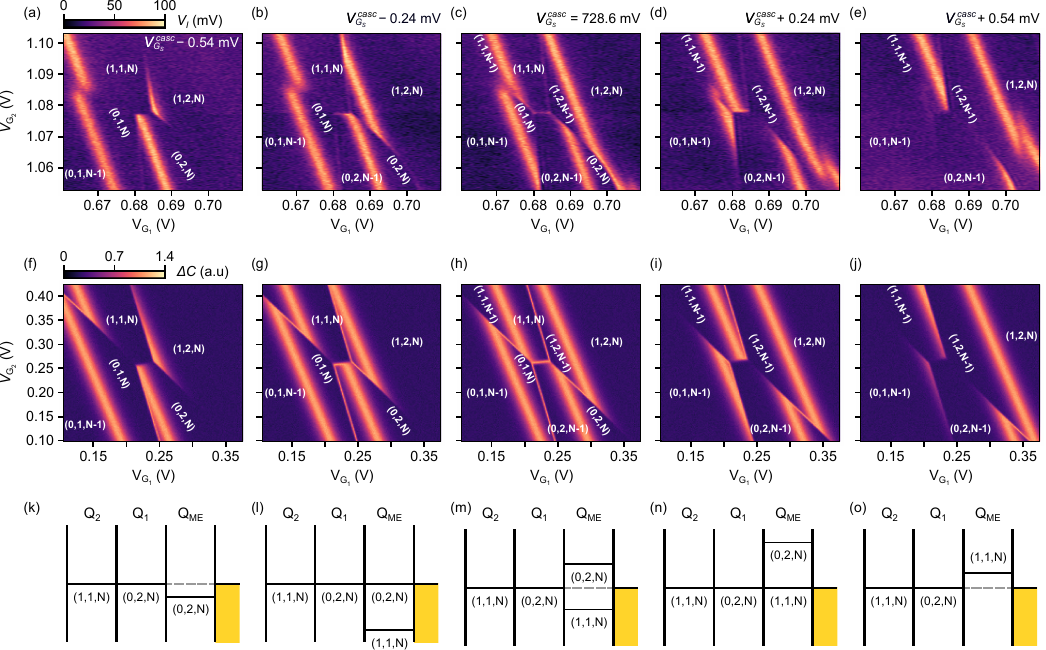}
    \def\figurename{\textbf{Extended Data Fig.}}
    \caption{\textbf{$|$ Tuning the radiofrequency driven electron cascade}. \textbf{a-e}, Charge stability diagrams as a function of the DQD gate voltages $V_{\text{G}_1}-V_{\text{G}_2}$. The multi-electron QD bias $V_{G_\text{S}}$ is varied in each panel, defined with respect to $V_{\text{G}_\text{S}}^\text{casc}$, the bias where the conditions for cascade are observed, shown in panel \textbf{c}. This is the bias utilized in the main text of the manuscript. Similar data for additional devices can be found in Supplementary Section 2. \textbf{f-j}, Simulated charge stability diagrams using the methods described in Supplementary Section 4. \textbf{k-o} Schematic depiction of the electrochemical levels for the corresponding bias conditions, arranged column-wise.}
    \label{FigSI:CSD}
\end{figure*}

\makeatletter
\setcounter{figure}{0} 
\renewcommand{\thefigure}{S\arabic{figure}}
\renewcommand{\figurename}{Supplementary Fig.}
\def\fnum@figure{\textbf{\figurename~\thefigure}}
\makeatother

\newcommand{\sifigref}[1]{\textbf{Supplementary Fig.~\ref{#1}}}
\begin{bibunit}

\renewcommand{\theequation}{{S\arabic{equation}}}
\renewcommand{\thesection}{{S\arabic{section}}}

\newlength{\supptoclabelwd}
\settowidth{\supptoclabelwd}{S10.\quad}

\newcommand{\suppentry}[3]{%
  \par\noindent
  \hangindent=\supptoclabelwd
  \hangafter=1
  \hyperref[#1]{%
    \makebox[\supptoclabelwd][l]{#2}%
    #3\dotfill \pageref*{#1}%
  }%
  \par\vspace{0.5em}%
}

\onecolumngrid

\clearpage
\phantomsection
\Large
\section*{Supplementary: Radiofrequency cascade readout of coupled spin qubits}
\large
\noindent
\suppentry{app:process}{S1.}{FABRICATION OF PROCESS CONTROL MONITORS}
\suppentry{app:cond}{S2.}{ELECTROSTATIC REQUIREMENTS FOR RF-DRIVEN ELECTRON CASCADE
READOUT}
\suppentry{app:amp}{S3.}{AMPLIFICATION FACTOR}
\suppentry{app:sim}{S4.}{CHARGE STABILITY DIAGRAM SIMULATION}
\suppentry{app:candle}{S5.}{MAGNETOSPECTROSCOPY SIMULATION}
\suppentry{app:QME}{S6.}{RADIOFREQUENCY ELECTRON CASCADE AS A FUNCTION OF $\rm Q_{\rm ME}$ OCCUPATION}
\suppentry{app:in-situ}{S7.}{IN-SITU DISPERSIVE PERFORMANCE}
\suppentry{app:ST0}{S8.}{SINGLET-TRIPLET QUBIT PERFORMANCE}
\suppentry{app:Echo}{S9.}{EXCHANGE ECHO PERFORMANCE}
\suppentry{app:Si-28}{S10.}{ESTIMATED IMPROVEMENT FROM ISOTOPICALLY ENRICHED $^{28}$Si}
\suppentry{app:supprefs}{}{SUPPLEMENTARY REFERENCES}

\normalsize
\newpage
\section{{Fabrication of process control monitors}}
\label{app:process}

\begin{figure}[h]
    \centering
    \includegraphics[width = 0.5\linewidth]{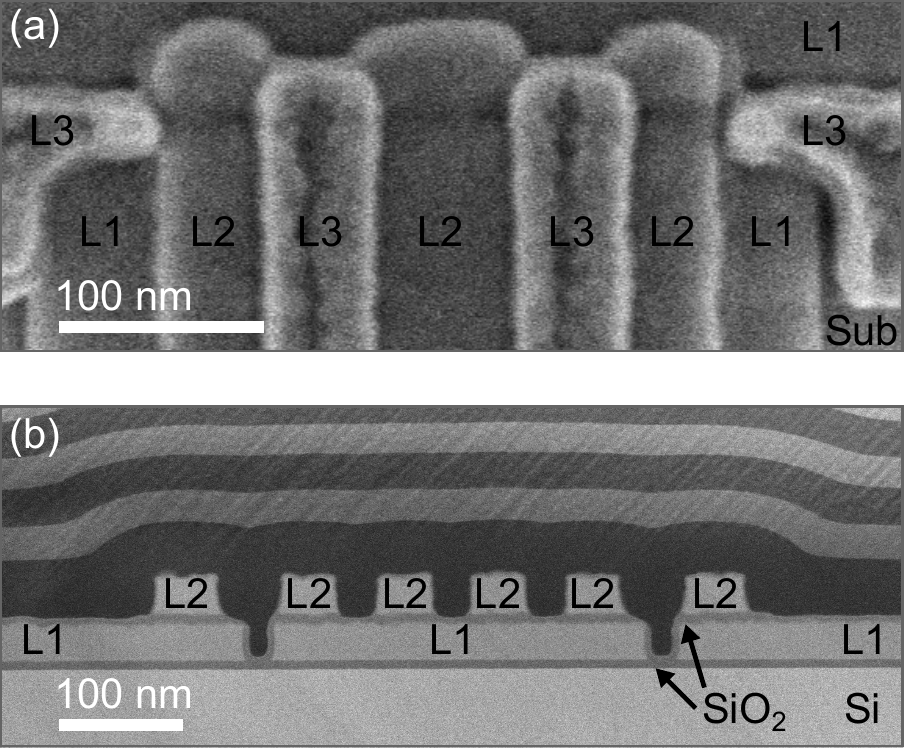}
    \caption{\textbf{$|$ Fabrication process control monitors (PCMs).} \textbf{a} Scanning electron microscopy image of a PCM linear quantum dot array, consisting of three overlapping gate layers: L1, L2 and L3 in order from bottom to top. \textbf{b} Cross-sectional transmission electron microscopy image of a separate PCM structure to \textbf{a}, with two overlapping gate layers, buried beneath several layers of oxide. The scale bar in each panel represents 100~nm.}
    \label{Fig:SEM_TEM}
\end{figure}

\noindent In this Supplementary Section, we illustrate the quality of the fabrication process via electron microscope images of process control monitors (PCMs), i.e non-functional test structures integrated into the semiconductor wafer used for morphological characterisation purposes. Figure~\ref{Fig:SEM_TEM}(a) shows a scanning electron microscopy (SEM) image of a PCM containing three gate layers and whose purpose is the creation of a linear quantum dot (QD) array. The PCM shares similarities with the array discussed in the main text. In particular, L1 gates confine the QD array to a line, L2 and L3 gates are used to control QDs, and L3 gates are used as reservoir accumulation gates. The PCM was fabricated with the same gate stack and used the same L2 gate length and pitch as the device in the main text (lateral L2 gates). The SEM image aims to illustrate features of the fabrication process such as critical dimensions, line edge roughness, and interlayer alignment accuracy. 

Additionally, in Fig.~\ref{Fig:SEM_TEM}(b), we show a cross-sectional transmission electron microscopy (TEM) image of a different PCM. The image illustrates the surface roughness of the Si-SiO$_2$ layer as well as the uniformity of the oxides between the gate layers. This TEM image only includes the first two gate layers.

\newpage
\section{{Electrostatic requirements for rf-driven electron cascade readout}}
\label{app:cond}

\noindent The conditions for rf-driven electron cascade are similar to those stated in a previous report on electron cascade with a proximal charge sensor \cite{vanDiepen:2021}. It can be understood by considering the electrochemical potentials $\mu_i$ for each QD Q$_i$, as well as the given charge configuration $(\text{N}_{\text{Q}_{2}}, \text{N}_{\text{Q}_{1}}, \text{N}_{\text{Q}_{\text{ME}}})$ where $\text{N}_{\text{Q}_{i}}$ refers to the number of charges in dot $\text{Q}_i$, as shown in Fig. 1 in the main text. For cascade to occur the DQD must be tuned to the $\text{Q}_2-\text{Q}_1$ inter-dot charge transition, such that 
\begin{equation} \label{eqn:cascade-1}
\mu_{\text{Q}_2}(1,1,\text{N}) = \mu_{\text{Q}_1}(0,2,\text{N}-1).
\end{equation}
Note that the opposite inter-dot charge transition $\mu_{\text{Q}_2}(2,0,\text{N}) = \mu_{\text{Q}_1}(1,1,\text{N})$ also satisfies this condition. In addition to this, the reservoir-adjacent dot ($\text{Q}_{\text{ME}}$) must be tuned such that
\begin{equation} \label{eqn:cascade-2}
    \mu_{\text{Q}_{\text{ME}}}(1,1,\text{N})<0<\mu_{\text{Q}_{\text{ME}}}(0,2,\text{N}),
\end{equation}
where the Fermi level of the reservoir is referenced to 0 and  
\begin{equation} \label{eqn:cascade-3}
    \Delta\mu_{\text{Q}_{\text{ME}}}= \mu_{\text{Q}_{\text{ME}}}(0,2,\text{N}) - \mu_{\text{Q}_{\text{ME}}}(1,1,\text{N})\gg 3.5k_\text{B}T,
\end{equation}
\noindent so the shift of the $\text{Q}_{\text{ME}}$ Coulomb oscillation due to the interdot charge transition is much larger than the Fermi broadening of the reservoir. Here $k_B$ is the Boltzmann constant and $T$ the temperature.

The rf mode of the casacde drive introduces an additional condition that relates to the applied rf modulation amplitude $V_{\text{rf}}$, 
\begin{equation} \label{eqn:cascade-4}
\Delta\mu_{\text{Q}_1-\text{Q}_2} < e V_{\text{rf}} \ll \Delta\mu_{\text{Q}_{\text{ME}}}
\end{equation}
where $\Delta\mu_{\text{Q}_1-\text{Q}_2} = \mu_{\text{Q}_1}(1,1,\text{N}) - \mu_{\text{Q}_2}(0,2,\text{N}-1)$. This ensures that the rf modulation only drives tunneling events between $\text{Q}_2$ and $\text{Q}_1$, without directly driving tunneling events between $\text{Q}_{\text{ME}}$ and the reservoir. 

Tuning the QD array into rf-driven electron cascade requires precise control of the electrochemical potentials of the QDs, to the order of tens of microvolts in the device presented in the main text (however, this bias range is extended to the order of a millivolt in measurements on additional devices, as noted in Supplementary Fig.~\ref{Fig:Ext-Casc}). The tuning procedure is most clearly demonstrated in the $V_{\text{G}_1}-V_{\text{G}_2}$ gate-voltage-space, as shown in Extended Data Fig.~1 and Supplementary Fig.~\ref{Fig:Ext-Casc}, where in each panel $V_{\text{G}_{\text{S}}}$ is varied, bringing the system into and out of cascade. In this configuration, the range of $V_{\text{G}_\text{S}}$ bias voltages satisfying the conditions for cascade are given by $V_{\text{G}_\text{S}}^{casc}=728.6\pm0.1~$mV (Extended Data Fig.~1), a relatively narrow range of voltages as compared to the $14~$mV addition voltage of $\text{Q}_{\text{ME}}$. 

Extended Data Figure 1~(a) and (e) shows two limiting cases, in which the $\mu_{\text{Q}_{\text{ME}}}(0,2,\text{N})$ or $\mu_{\text{Q}_{\text{ME}}}(1,1,\text{N})$ potentials do not meet the condition set by Eq.~(\ref{eqn:cascade-2}), i.e. the levels are below or above the Fermi level of the reservoir, respectively. Similar cases are observed in panels (b) and (d), where $\mu_{\text{Q}_{\text{ME}}}(0,2,\text{N})$ or $\mu_{\text{Q}_{\text{ME}}}(1,1,\text{N})$ now just align with the Fermi level in the reservoir but are within its Fermi broadening. In panel (b), the $\text{Q}_{\text{ME}}$ transition is present in the $(0,2,\text{N-}1)-(0,2,\text{N})$ occupation regime, but absent in the $(1,1,\text{N-}1)-(1,1,\text{N})$ regime. Likewise, in panel (d), the $\text{Q}_{\text{ME}}$ transition is present in the $(1,1,\text{N-}1)-(1,1,\text{N})$ regime but absent in the $(0,2,\text{N-}1)-(0,2,\text{N})$ regime. The contrast in signal between the different charge occupations shown in Extended Data Fig. 1(b) and (d) is well-suited to standard charge sensing. It is only in panel (c), when the sequential tunneling event occurs, the cascade conditions are met and we observe the enhanced intensity of the interdot charge transition. 

\setcounter{figure}{1} 
\begin{figure}[h]
    \centering
    \includegraphics[width = \linewidth]{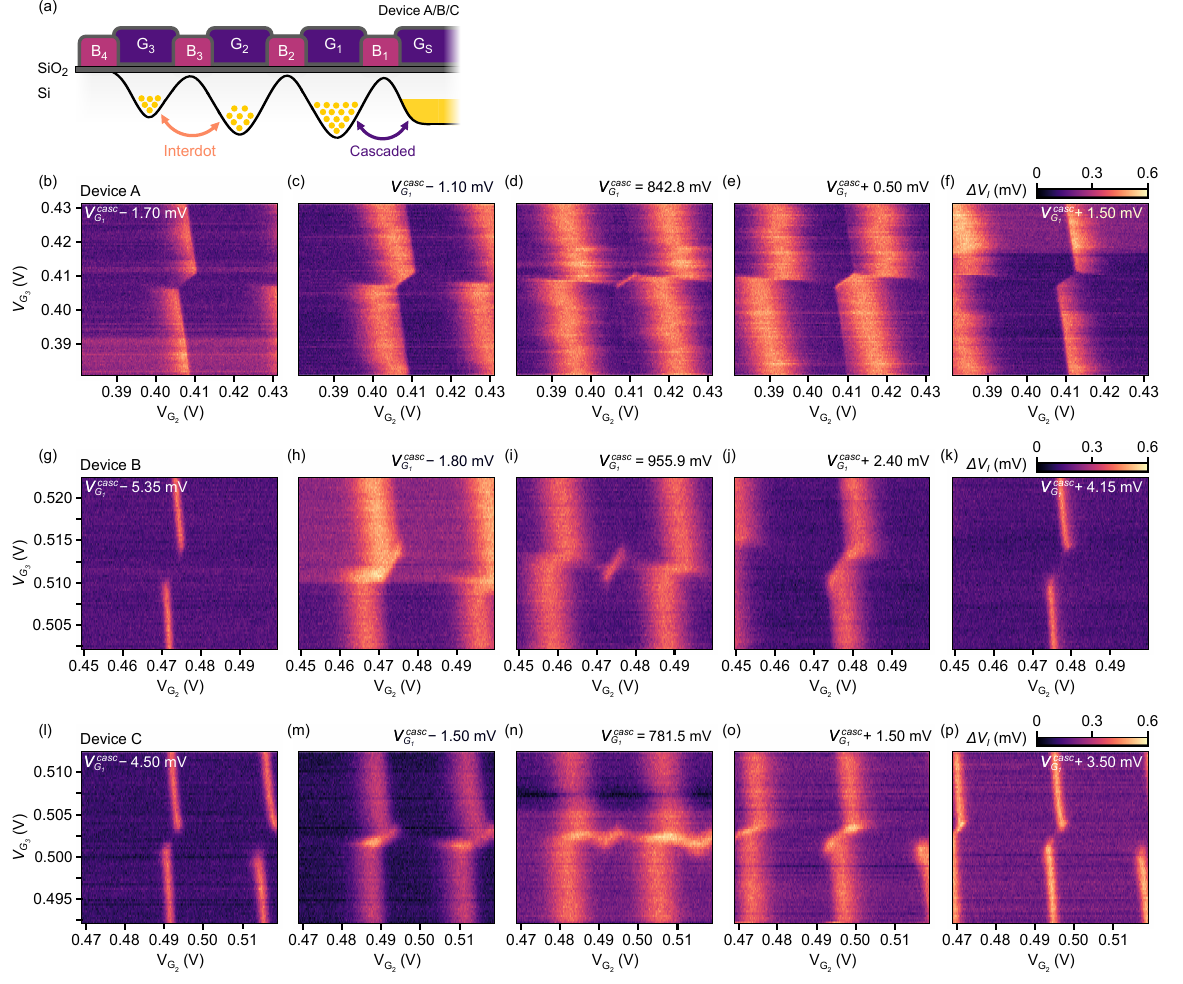}
    \caption{\textbf{$|$ Radio-frequency cascade reproduced across multiple devices.} \textbf{a} Cross-section schematic of the three quantum dot arrays, A, B, and C, where each array has varied gate dimensions. The exact number of electrons occupying each QD is unknown. Each row depicts each of the devices being tuned around the cascade regime (centre column) for: \textbf{b-f} Device A $(V_{\rm G_S}^{\rm casc} = 842.8\pm1.0~{\rm mV})$, \textbf{h-l} Device B $(V_{\rm G_S}^{\rm casc} = 955.9\pm1.2~{\rm mV})$, \textbf{m-p} Device C $(V_{\rm G_S}^{\rm casc} = 781.5\pm1.5~{\rm mV})$.}
    \label{Fig:Ext-Casc}
\end{figure}

In panels (f-j), we present matching radio-frequency simulations of the triple QD system that highlight the enhanced intensity of the interdot charge transition (see Section.~\ref{app:sim}). Further, we supplement the explanation with schematics of the electrochemical levels in each $V_{\text{G}_\text{S}}$ conditions in panels (k-o). 

Supplementary Figure~\ref{Fig:Ext-Casc}, shows the rf-cascade tuning process across three additional devices (A,B, and C), where each device is a linear array of QDs similar to the device discussed in the main text, but with the addition of inter-dot barrier gates. The barriers enable finer control of inter-dot coupling and broaden the cascade window from tens of microvolts to the order of a millivolt. Each row illustrates the same tuning sequence as Extended Data Fig.~1, as $V_{G_S}$ is increased $\mu_{\rm Q_{ME}}$ is tuned with respect to the Fermi level of the electron reservoir, in and out of the cascade regime. The cascade window for each device is defined by $V_{\rm G_S}^{\rm A,casc} = 842.8\pm1.0~{\rm mV}$, $V_{\rm G_S}^{\rm B,casc} = 955.9\pm1.2~{\rm mV}$, and $V_{\rm G_S}^{\rm C,casc} = 781.5\pm1.5~{\rm mV}$ respectively.

\newpage
\section{{Amplification Factor}} \label{app:amp} \noindent
Here, we obtain the expression for the signal amplification factor generated by the cascade process, i.e. Eq.~(1) in the main text. To determine the amplification factor, we consider two different charge movement events as seen from the electrode connected to the resonator, in this case, the electron reservoir (R):
\begin{enumerate}
    \item \textbf{In-situ dispersive readout} involving solely a charge transition between Q$_1$ and Q$_2$.
    \item \textbf{Cascade readout} involving (1) plus the cascaded charge transition between Q$_\text{ME}$ and the reservoir.  
\end{enumerate}

\setcounter{figure}{2} 
\begin{figure}[h]
    \centering
    \includegraphics[width = 0.5\linewidth]{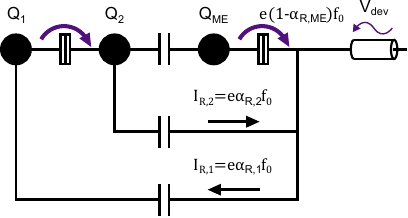}
    \caption{\textbf{$|$ Schematic circuit depicting alternating current induced by the radio-frequency driven electron cascade.} Two tunneling processes are depicted between tunnel coupled (1) quantum dots Q$_1$ and Q$_2$ and (2) Q$_{\rm ME}$ and the reservoir. The induced alternating current at the reservoir from each of these tunneling events is given by $I_{\rm R,i} = e \alpha_{R,i} f_0$ where $i = 1,2,{\rm ME}$ and $\alpha_{\rm R,i}$ are the corresponding lever arms, with rf drive frequency $f_0$. The rf modulation applied to the system is represented by $V_{\rm dev}$, the device voltage.}
    \label{Fig:CascCirc}
\end{figure}

To obtain the amplification factor, we first consider the expressions for the signal-to-noise ratio for case (1)~\cite{Vigneau:2023}. In particular, we assume that the system is in the low frequency limit $hf_\text{rf}\ll \Delta_\text{c}$ (where $f_\text{rf}$ is the frequency of the resonator and $\Delta_\text{c}=2t_c$), the small signal regime $Q_\text{L}\Delta C_\text{Q}/(2C_\text{tot})\ll 1$ (where $Q_\text{L}$ is the loaded quality factor of the resonator, $\Delta C_\text{Q}$ is the change in quantum capacitance of the system and $C_\text{tot}$ the total capacitance of the system) and the large excitation regime, $\alpha_{21} e V_\text{dev} \gg \Delta_\text{c}$ to ensure an electron tunnels every half cycle of the rf excitation. Here $\alpha_{21}=\alpha_\text{R,2}-\alpha_\text{R,1}$ is the interdot lever arm as seen from the reservoir and $V_\text{dev}$ is the amplitude of the oscillatory voltage arriving at the reservoir. In this case, the SNR is 
\begin{equation}\label{eq:ict}
    \text{SNR}_\mathrm{21} \propto \frac{(\alpha_{21}e)^2}{k_\text{B}T_\text{n}}Q_0Z_\text{r}f_\text{rf}^2= \frac{I_\text{R,21}^2 R}{k_\text{B}T_\text{n}},
\end{equation}
\noindent where $T_\text{n}$ is the noise temperature of the system, $Q_0=R\sqrt{C_\text{p}/L}$ is the internal quality factor of the resonator, $Z_\text{r}=\sqrt{L/C_\text{p}}$ is the resonator impedance, $R$ is the resistance (in parallel with $L$ and $C_\text{p}$) representing the losses in the resonator, and $I_\text{R,21}$ is the AC current amplitude produced at the reservoir by the oscillatory charge motion between Q$_1$ and Q$_2$. Note that $\alpha_{21}ef_\text{rf}=I_\text{R,21}$ is the induced current at the reservoir due to the cyclic interdot tunneling. 

We now perform the same calculation for the charge transition between the multi-electron QD $Q_\text{ME}$ and the reservoir $R$, a dot-to-reservoir transition~\cite{Oakes:2023}. Again, considering the small signal regime, the large excitation regime, $\alpha_\text{R,ME} e V_\text{dev} \gg k_\text{B}T$ (where $\alpha_\text{R,ME}$ is the lever arm from the reservoir to Q$_\text{ME}$ and $T$ is the electron temperature) and considering additionally the fast tunneling regime $\gamma\gg f_\text{rf}$ (where $\gamma$ is the tunnel rate between Q$_\text{ME}$ and R), we obtain
\begin{equation}\label{eq:seb}
    \text{SNR}_\mathrm{ME} \propto \frac{(1-\alpha_\text{R,ME})^2e^2}{k_\text{B}T_\text{n}}Q_0Z_\text{r}f_\text{rf}^2= \frac{I_\text{R,ME}^2 R}{k_\text{B}T_\text{n}}.
\end{equation}
Here, the resonator is coupled to $Q_\text{ME}$ via a tunnel barrier, rather than a capacitively coupled gate, and hence the induced charge is $(1-\alpha_\text{R,ME})e$. 

After this analysis, we highlight a critical result: the SNR is proportional to the square of the AC current amplitude produced by the relevant process, whereas the other two parameters, $R$ and $T_\text{n}$, are independent of the charge transfer process, particularly in the typical regime where the noise temperature is determined by the first amplifying stage.

From this we can determine an expression for the amplification factor by considering the two tunneling processes presented in Supplementary Fig.~\ref{Fig:CascCirc}. In process (1) the alternating current generated by cyclical tunneling from Q$_1$ to Q$_2$ is,
\begin{equation}\label{eq:Iict}
    I_\text{R,21}=\alpha_{21}ef_\text{rf}=(\alpha_\text{R,2}-\alpha_\text{R,1})ef_\text{rf}.
\end{equation}
In process (2), in addition to the current produced by process (1), we have the current produced by the cascade process adding to a total of,
\begin{equation}\label{eq:Icasc}
    I_\text{cascade} \approx I_\text{R,21}+I_\text{R,ME}=(\alpha_\text{R,2}-\alpha_\text{R,1})ef_\text{rf}+(1-\alpha_\text{R,ME})ef_\text{rf}. 
\end{equation}
The amplification factor $A$ is then given by the ratio between the two processes,
\begin{equation}\label{eq:Icasc}
   A =  \frac{I_\text{cascade}}{I_\text{R,21}}=1+\frac{1-\alpha_\text{R,ME}}{\alpha_\text{R,2}-\alpha_\text{R,1}}. 
\end{equation}

\section{{Charge Stability Diagram Simulation}} \label{app:sim}

\noindent To simulate the cascade phenomena we observe in the main text, we calculate the charge stability diagram for a specific voltage configuration using the Constant Interaction Model \cite{vanderWiel:2002}. The energies in this model are defined as
\begin{equation}
E = \frac{1}{2} \vec{V}^{T} \mathbf{C_{cc}^{-1}} \vec{V}
\end{equation}
where $\mathbf{C_{cc}}$ is the capacitance matrix for the QDs, containing the mutual capacitance between each pair

\begin{equation}
C_{cc} = \begin{pmatrix}
35.000 & -4.882 & -1.886 \\
-4.882 & 27.936 & -0.402 \\
-1.886 & -0.402 & 42.048
\end{pmatrix}
\end{equation}
where each element is given in aF and $\vec{V} = e (\mathbf{C_{cv}} \vec{V_G} - |e|\vec{N})$. Additionally we define $\mathbf{C_{cv}}$ to represent the capacitance matrix governing the interactions between the gate and charges, containing the capacitance between each QD and the corresponding gates,
\begin{equation}
C_{cv} = \begin{pmatrix}
0.898 & 0.146 & 0.000 \\
0.537 & 0.453 & 0.000 \\
0.063 & 0.020 & 2.016
\end{pmatrix}
\end{equation}
where each element is given in aF. The number of charges on each dot are given by $\vec{N}$, and $\vec{V_G}$ denotes the applied gate voltages. Throughout the simulations, we use natural units and set the charge of the electron  $e = 1$. We note that the ratio between capacitances in the matrices defined here are found to be in qualitative agreement with the data in Extended Data Fig.~1(a-e), however the magnitude of these capacitances have not been verified experimentally. 

Without loss of generality, we assume that an $M$-quantum dot array can exist in a state from a set of $L$ Fock states, denoted as $\mathcal{F} = \left\{ \Lambda_j = (\lambda_{1,j}, \ldots, \lambda_{M,j}) \mid \lambda_{i,j} \in \mathbb{Z}^+, \, \forall j = 1, 2, \ldots, L \right\}$. Here $\lambda_{i,j}$ represents the occupancy number of QD $i$ in Fock state $\Lambda_j$. 

In radio-frequency reflectometry, the measured signal is directly proportional to the change in capacitance of the system, this can be described mathematically as
\begin{equation}
    \Delta C_j = \frac{d Q_T}{d V_j},
    \label{eqn:C_j}
\end{equation}
here $Q_T$ denotes the total charge of the system and $V_j$ represents the $j^{th}$ gate. We take inspiration from \cite{Mizuta:2017} and rewrite the change in capacitance as measured from gate $V_j$ to be
\begin{equation}
    \Delta C_{j} = \sum_{i=1}^M C_{i, j} = e \sum_{i=1}^M  \alpha_{i, j} \frac{d\langle n_i \rangle}{dV_j} .
    \label{eqn:BIG}
\end{equation}
In this context, \(C_{i,j}\) represents the capacitance felt from dot \(i\) by gate \(j\). The average occupancy of dot \(i\) is denoted \(\langle n_i \rangle\). The lever arm matrix is defined as \(\alpha = \mathbf{C_{cc}^{-1}}\mathbf{C_{cv}}\), which links gate-induced potential changes to the charge states of the QDs. To calculate $\langle n_i \rangle$, we iterate through each Fock state in $\mathcal{F}$ and compute its corresponding probability. Subsequently, a weighted sum of the occupation numbers of each Fock state at position $i$ is performed, expressed as
\begin{equation}
     \langle n_i \rangle = \sum_{k = 1}^{L} \lambda_{i, k} \cdot P_{k},
     \label{eqn:n_i2}
\end{equation}
where $P_{k}$ represents the probability of the QD array being in the Fock state $\Lambda_k$. We assume a Boltzmann distribution and write the probability accordingly;
\begin{equation}
    P_{k} = \frac{1}{Z} \exp \left( -\frac{\epsilon_{k}}{k_bT} \right) = \frac{\exp \left( -\frac{\epsilon_{k}}{k_bT} \right)}{\sum_{l=1}^L \exp \left( -\frac{\epsilon_{l}}{k_bT} \right)}
\end{equation}
where \( Z \) represents the partition function, and \( \epsilon_{k} \) represents the energy required for QD array to be in the Fock state $\Lambda_k$. Substituting this into equation \ref{eqn:n_i2}, we obtain;
\begin{equation}
     \langle n_i \rangle = \sum_{k = 1}^{L} \lambda_{i, k} \cdot  \frac{\exp \left( -\frac{\epsilon_{k}}{k_bT} \right)}{\sum_{l=1}^L \exp \left( -\frac{\epsilon_{l}}{k_bT} \right)}
     \label{eqn:n_i3}
\end{equation}
This term can be further substituted into equation \ref{eqn:BIG}, to obtain;
\begin{align}
    \Delta C_{j, \text{tot}} &= \sum_{i=1}^M C_{i,j} \\
    &= e \sum_{i} \alpha_{i,j} \frac{d}{dV_j} 
    \left(
    \sum_{k = 1}^{L} \lambda_{i, k} \cdot  \frac{\exp \left( -\frac{\epsilon_{k}}{k_bT} \right)}{\sum_{l=1}^L \exp \left( -\frac{\epsilon_{l}}{k_bT} \right)}
    \right)
    \label{eqn:P_eqn_old}
\end{align}
We use the above formalism to simulate the charge stability diagrams in the $V_{\text{G}_1}-V_{\text{G}_2}$ gate-voltage space. This can be seen in Extended Data Fig. 1(f-j), where different voltage configurations were applied to the multi-electron dot via $V_{\text{G}_\text{S}}$. 

\newpage
\section{{Magnetospectroscopy simulation}} \label{app:candle}

\noindent Here, we describe the simulations of the magnetospectroscopy map in the inset of Fig.~1(c) in the main text. that allow us to estimate both the tunnel coupling $t_\text{c}$ and the electron temperature $T_\text{e}$. We utilize the simplified Hamiltonian:
\begin{equation}\label{Eq:Hamil}
	H=\frac{1}{2}
	\begin{pmatrix}
		\varepsilon & \Delta_\text{c} & 0 & 0 & 0 \\ 
		\Delta_\text{c} & -\varepsilon & 0 & 0 & 0\\
		0 & 0 & -\varepsilon-\hat{B} & 0 & 0\\
		0 & 0 & 0 & -\varepsilon & 0 \\
		0 & 0 & 0 & 0 & -\varepsilon+\hat{B}
	\end{pmatrix},
\end{equation}
\noindent where $\hat{B}=2g\mu_\text{B}B$, $g$ is the electron g-factor (which we approximate to 2 for both QDs). Then, we calculate the quantum capacitance of the system, $C_\text{Q}$, given by,
\begin{equation}
	C_\text{Q}=-\sum_i (e\alpha)^2\frac{\partial^2 E_i}{\partial\varepsilon^2}P^\text{th}_i.
\end{equation}
\noindent where $E_i$ are the eigenenergies of the above Hamiltonian and $P^\text{th}_i$ is the thermal probability of the state $i$,
\begin{equation}
	P^\text{th}_i=\exp(-E_i/k_\text{B}T_\text{e})/Z.
\end{equation}

Here, $T_\text{e}$ is the DQD temperature and $Z$ is the partition function over all states~\cite{Lundberg:2020}. We plot the results of the simulations in Supplementary Fig.~\ref{Fig:SB} where, in panel (a) and (b), we show the energy spectrum of the system as a function of detuning for $B=0.5$~T and 0~T, respectively, and in panel (c), we plot the normalised quantum capacitance of the systems as a function of $\varepsilon$ and $B$. We find that the best match between the data and simulations occurs when $t_\text{c}=2.4$~GHz and $T_\text{e}=50$~mK. 

\setcounter{figure}{3} 
\begin{figure}[h]
    \includegraphics{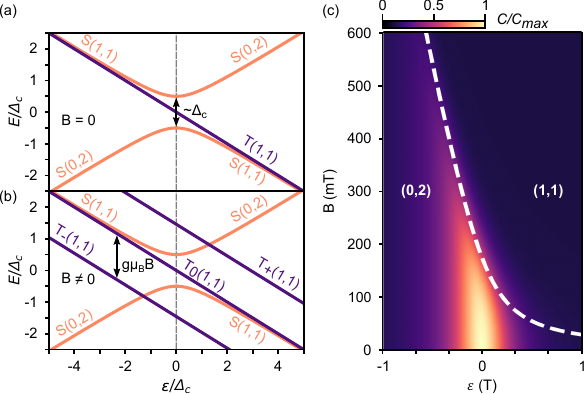}
    \caption{\textbf{$|$ Magneto-spectroscopy simulation of the (0,2)-(1,1) inter-dot charge transition}. Eigenenergies of the two-electron spin states, singlet $\ket{\text{S}}$ and triplet $\ket{\text{T}}$ states as a function of detuning $\varepsilon$ normalised by tunnel coupling $\Delta = 2 t_c$. For \textbf{a}, applied magnetic field $B=0~$mT and \textbf{b} $B=500~$mT. \textbf{c} Normalised quantum capacitance simulation versus detuning and magnetic field, with the degeneracy point of the $\ket{\text{S}}$-$\ket{\text{T}_-}$ crossing overlaid (white dashed line).}
    \label{Fig:SB}
\end{figure}

\newpage
\section{{Radio-frequency electron cascade as a function of Q$_{\rm ME}$ occupation}} \label{app:QME}
\noindent
We observe that the occupancy of Q$_{\rm ME}$, does not have a significant effect on the cascade signal. Supplementary Figure~\ref{Fig:ME_Occupation} shows the charge stability diagrams measured at different occupations N$_{\rm ME}$, in each case we see the cascaded interdot charge transition (Fig~S\ref{Fig:ME_Occupation}(a-b)). Additionally, we observe Pauli spin blockade in each configuration, as evident from the magnetospectroscopy measurements in Fig.~S\ref{Fig:ME_Occupation}(c-d), where the disappearance of signal for increasing $|B|$ indicates the $\ket{\rm S}$ is no longer in the ground state at $\varepsilon = 0$.

\setcounter{figure}{4} 
\begin{figure}[h]
    \centering
    \includegraphics[width = 0.55\linewidth]{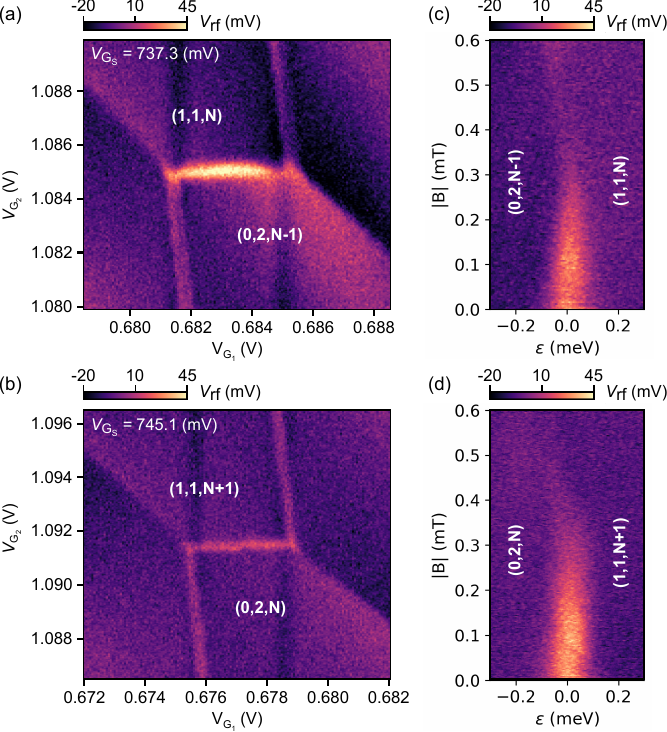}
    \caption{\textbf{$|$ Radio-frequency driven electron cascade at different Q$_{\rm ME}$ occupations.} Charge stability diagrams \textbf{(a-b)} and magnetospectroscopy \textbf{(c-d)} for Q$_{\rm ME}$ transitions, $(\text{N}_{\rm ME}-1)\rightarrow(\text{N}_{\rm ME})$ and $(\text{N}_{\rm ME})\rightarrow(\text{N}_{\rm ME}+1)$, respectively. *We note that the cascade observed in the charge stability diagram in panel (\textbf{b}) is not optimally tuned, but it is in the corresponding magnetospectroscopy measurement in panel \textbf{(d)}. Furthermore, the dark lines observed in the charge stability diagrams, at $V_{\rm G_2} = 0.682~$V and $0.685~$V in panel \textbf{(a)},are the result of the AC-coupled measurement mode.}
    \label{Fig:ME_Occupation}
\end{figure}

\newpage
\section{{In-situ dispersive readout performance}} \label{app:in-situ}

\begin{table}[h] 
    \large
    \begin{center}
    \begin{tabular}{r c c c c c}
    \hline
    Reference & Platform & Resonator & $\tau_{\rm int}~ (\upmu\text{s})$ & $\mathcal{F}_r~(\%)$ & $T_1~(\upmu s)$  \\
    \hline
    Pakkiam \textit{et al.} 2018~\cite{Pakkiam18} & Donor in Si & Off-chip SC & 300 & 82.9 & 620 \\
    West \textit{et al.} 2019~\cite{West:2019} & Planar $\text{Si/SiO}_2$ & SMD & 2000 & 73.3 & 4500 \\
    Zheng \textit{et al.} 2019~\cite{Zheng:2019} & $\text{Si/SiGe}$ & On-chip SC & 6 & 98.0 & 159 \\
    Chittock-Wood \textit{et al.} 2025 & Planar $\text{Si/SiO}_2$ & Off-chip SC & 8 & 67.0 & 24 \\
    \hline
    \end{tabular}
    \end{center}
    \def\tablename{\textbf{Supplementary Table}}
    \caption{\textbf{In-situ dispersive readout performance across different silicon technology platforms.} Here resonators are categorised into three-types, on- and off-chip superconducting (SC) resonators and those constructed using surface mounted (SMD) circuit components.}
    \label{tab:fidelity}
\end{table}

\section{{Singlet-Triplet Qubit Performance}} \label{app:ST0}
\begin{table}[h] 
    \large
    \begin{center}
    \begin{tabular}{r c c c}
    \hline
    Reference & Platform & Q & $\tau_{\rm gate}$ (ns) \\
    \hline
    Dial \textit{et al.} 2013~\cite{Dial:2013} & \text{GaAs/AlGaAs} & 6 & 0.6 \\
    Wu \textit{et al.} 2014~\cite{Wu:2014} & \text{Si/SiGe} & 1.5 & 36 \\
    Higginbotham \textit{et al.} 2014~\cite{Higginbotham14} & \text{GaAs/AlGaAs} & 15 & 0.3 \\
    Jang \textit{et al.} 2020~\cite{Jang20} & \text{GaAs/AlGaAs} & 26 & 1.1 \\
    Jirovec \textit{et al.} 2021~\cite{Jirovec:2021} & \text{Ge/SiGe} & 50 & 8 \\
    Jock \textit{et al.} 2018~\cite{Jock:2018} & $^{28}\text{Si/SiO}_2$ & 2 & 250 \\
    Connors \textit{et al.} 2022~\cite{Connors22} & $\text{Si/SiGe}$ & 5.3 & 17 \\
    Chittock-Wood \textit{et al.} 2025 & $\text{Si/SiO}_2$ & $\geq$10 & 22 \\
    \hline
    \end{tabular}
    \end{center}
    \def\tablename{\textbf{Supplementary Table}}
    \caption{\textbf{Singlet-triplet qubit quality factors and gate times across different semiconductor platforms,} where $Q = T_2^* \Omega $ is the qubit quality factor calculated using the qubit frequency $\Omega$ and dephasing time $T_2^*$, and $\tau_{\rm gate}=1/2\Omega$ is the qubit gate time.}
    \label{tab:Q-factor}
\end{table}

\section{{Exchange Echo Performance}} \label{app:Echo}

\begin{table}[h] 
    \large
    \begin{center}
    \begin{tabular}{r c c c c c c c}
    \hline
    Reference & Platform & $\delta\varepsilon_{\rm rms}~(\upmu\text{eV})$ & $\Delta E_{\rm z,rms}~({\rm neV})$ & $T_2^{*}~ (\upmu\text{s})$ & $T_2^{\text{echo}}~(\upmu\text{s})$ & $T_{\rm2, mag}^*~(\upmu\text{s})$ \\
    \hline
    Jock \textit{et al.} 2018~\cite{Jock:2018} & $^{28}\text{Si/SiO}_2$ & 2.0 & $\leq$ 0.2$^{a}$ & 1.00 & 8.4 & $\geq$20 \\
    Connors \textit{et al.} 2022~\cite{Connors22} & $\text{Si/SiGe}$ & 2.7 & $\leq$1.3$^{b}$ & 0.10 & 0.4 & $\geq$9 \\
    Chittock-Wood \textit{et al.} 2025 & $\text{Si/SiO}_2$ & 5.4 & 3.4 & 0.04 & 0.4 & 3 \\
    \hline
    \end{tabular}
    \end{center}
    \def\tablename{\textbf{Supplementary Table}}
    \caption{\textbf{Exchange echo reports in silicon quantum devices,} where a singlet-triplet qubit is configured to dephase under exchange $J$ and is refocused via the difference in Zeeman energy between the two dots $\Delta E_z$~\cite{Dial:2013} (in a DQD). Noise parameters $\delta\varepsilon_{\rm rms}$ and $\delta \Delta E_{\rm z, rms}$ describe rms fluctuations in detuning $\varepsilon$ and $\Delta E_z$, respectively. Note that here, the dephasing time $T_2^*$ listed does not represent the longest $T_2^*$ achieved in each reference, but is the reported value at the detuning $\varepsilon_{\text{echo}}$ where the echo was implemented. The echo time is reported as $T_2^{\rm echo}$, and this is contrasted against the magnetic limited dephasing time $T_{\rm 2,mag}^* = \lim_{\delta\varepsilon_{\rm rms} \to 0} T_2^*$, $^{a-b}$determined by $\sqrt{2}\hbar/\lim_{\varepsilon \to \infty}T_2^*$ in $^{a}$Fig.~4(d)~\cite{Jock:2018} and $^{b}$Fig.~2(d)~\cite{Connors22}.}
    \label{tab:echo}
\end{table}

\newpage
\section{{Estimated improvement from isotopically enriched $^{28}$S}i} \label{app:Si-28}

\noindent
The random Overhauser fluctuations in this effective magnetic field are then described by, 
\begin{equation} \label{eqn:HFI}
    \delta\mathcal{A} = \frac{\mathcal{A}}{\sqrt{N_{\text{nuc}}}} = \frac{\hbar}{T_{2,m}^*},
\end{equation}
it then follows that the rms magnetic noise calculated in the main text is equivalent to $\delta\Delta E_{z,rms} = \sqrt{2} {\delta \mathcal{A}}$, we therefore find $\delta \mathcal{A} = 2.4\pm0.3~$neV $(21 \pm 3~\upmu\text{T})$. The corresponding number of $^{29}$Si nuclei that an electron interacts can then be obtained by rearranging eq.~(\ref{eqn:HFI}) such that,
\begin{equation}
    N_{\text{nuc}} = \left(\frac{\hbar}{\mathcal{A}_0 T_{2,m}^*}\right)^2
\end{equation}
where $\mathcal{A}_0 = \mathcal{A}/N_{\text{nuc}}=0.043~$neV is the HFI energy per nuclei for Si~\cite{Assali11}. We find $N_{\text{nuc}} = 3100\pm800$ for the $T_{2,m}^*$ measured in the main text, which corresponds to the $4.67\%$ natural abundance of $^{29}$Si~\cite{Audi1997}. Furthermore, we can estimate the volume of the QD from,
\begin{equation} \label{eqn:VQD}
    V_{\text{QD}} = \frac{N_{\text{nuc}}}{f_{^{29}\text{Si}}\rho_{\text{atom}}},
\end{equation}
where $\rho_{\text{atom}}=4.99\times10^{28}~\text{m}^{-3}$ is the atomic density of Si, and $f_{^{29}\text{Si}}$ is the fraction of $^{29}$Si nuclei per m$^{-3}$, indicating a QD volume of $V_{\text{QD}} = 1300\pm300~\text{nm}^3$. Then to calculate $T_{2,m}^*$ for 800 ppm $^{29}$Si, we substitute $f_{^{29}\text{Si}} = 0.08\% \times 0.0467\%$ into eq.~(\ref{eqn:VQD}) and re-arrange to find $N_{\text{nuc}}^{\text{800ppm}} = 2\pm1$. Finally, substituting this result in eq.~(\ref{eqn:HFI}) we find $T_{2,m}^* = \hbar/\mathcal{A}_0 \sqrt{N_{\text{nuc}}^{\text{800ppm}}} = 9.8\pm0.8~\upmu$s, corresponding to an improvement of an order of magnitude. This is consistent with studies of similar two-electron systems defined in 800 ppm isotopically purified planar $^{28}$Si/SiO$_2$ devices, where $T_2^* \approx 1-3~\upmu$s~\cite{Harvey-Collard17, Fogarty2018, Jock:2018, Jock:2022}

\setcounter{figure}{5} 
\begin{figure}[h]
    \centering
    \includegraphics[width = 0.95\linewidth]{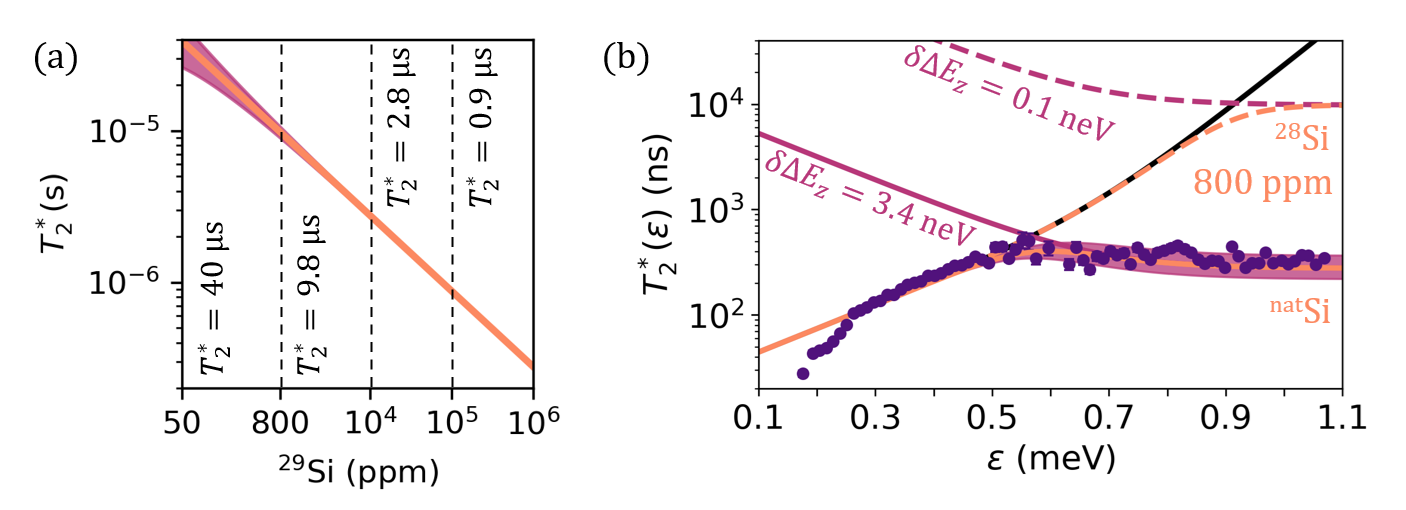}
    \caption{\textbf{$|$ Expected extension to $T_2^*$ from isotopic purification.} \textbf{a}, Extrapolated dephasing time $T_2^*$ as a function of $^{29}$Si isotopic abundance in the silicon substrate, in units of parts per million (ppm). \textbf{b}, Dephasing time as a function of detuning $\varepsilon$, with dataset for a natural silicon substrate and fits using Eq.~(4) (solid orange line), with magnetic noise and electron noise contributions to $T_2^*(\varepsilon)$ depicted by solid light purple and black lines respectively. The dashed lines show the corresponding estimate for a 800 ppm isotopically enriched silicon substrate. Shaded area shows the propagated error from the $\pm1$ standard deviation of fit parameters.}
    \label{Fig:Si-28}
\end{figure}

\newpage
\label{app:supprefs}
\putbib
\end{bibunit}

\end{document}